\documentclass[pra,twocolumn,byrevtex,showpacs]{revtex4}

\usepackage{graphicx}
\usepackage{dcolumn}
\usepackage{amsmath}
\usepackage{epsfig}
\usepackage{bm}
\usepackage{amssymb}

\usepackage[dvips]{hyperref}
\hypersetup{colorlinks=true,linkcolor=blue,citecolor=blue,urlcolor=black}

\begin{document}

\title{Remote state preparation of spatial qubits}

\date{\today}
\author{M. A. Sol\'is-Prosser}
\author{L. Neves}
\email{leonardo.neves@cefop.udec.cl} 

\affiliation{Center for Optics and Photonics, Universidad de Concepci\'on, Casilla 4016, Concepci\'on, Chile}
\affiliation{Departamento de F\'isica, Universidad de Concepci\'on, Casilla 160-C, Concepci\'on, Chile}



\begin{abstract}  

We study the quantum communication protocol of remote state preparation (RSP) for pure states of qubits encoded in single photons transmitted through a double slit, the so-called spatial qubits. Two measurement strategies that one can adopt to remotely prepare the states are discussed. The first strategy is the well-known spatial postselection, where a single-pixel detector measures the transverse position of the photon between the focal and the image plane of a lens. The second strategy, proposed by ourselves, is a generalized measurement divided into two steps: the implementation of a two-outcome positive operator-valued measurement (POVM) followed by the spatial postselection at the focal plane of the lens by a two-pixel detector in each output of the POVM. In both cases we analyze the effects of the finite spatial resolution of the detectors over three figures of merit of the protocol, namely, the probability of preparation, the fidelity and purity of the remotely prepared states. It is shown that our strategy improves these figures compared with spatial postselection, at the expense of increasing the classical communication cost as well as the required experimental resources. In addition, we present a modified version of our strategy for RSP of spatial qudits which is able to prepare arbitrary pure states, unlike spatial postselection alone. We expect that our study may also be extended for RSP of the angular spectrum of a single-photon field  as an alternative for quantum teleportation which requires very inefficient nonlinear interactions. 

\end{abstract}

\maketitle

\section{Introduction}

Remote state preparation (RSP) \cite{Lo00,Pati00,Bennett01} is a quantum communication protocol similar to quantum teleportation (QT) \cite{Bennett93}, regarding their ultimate goals: a sender, Alice, wants to transmit to a receiver, Bob, the quantum state of a qubit without sending physically the information carrier. In both protocols Alice and Bob need to share entanglement and the state of Bob's system is remotely prepared, after Alice has communicated classically to him the outcome of a measurement on her system. However, unlike QT, in RSP Alice has full knowledge of the state to be sent. For this reason, while in QT the consumption of one maximally entangled state (ebit) and two bits of classical communication (cbits) are both necessary and sufficient for faithfully transmitting the state, in RSP it is possible to trade off between these two resources \cite{Lo00,Pati00,Bennett01,Bennett05,Ye04}. For instance, a qubit lying in a given great circle of the Bloch sphere can be remotely prepared  with 1 cbit from Alice to Bob, if they share 1 ebit of entanglement \cite{Pati00}. Also, it is possible that Alice and Bob share a nonmaximally entangled state (i.e, less than 1 ebit) and she can remotely prepare his qubit either in a state within specific ensembles using 1 cbit or arbitrary pure states using finite (more than 1) cbits \cite{Ye04}.

Another important aspect that distinguishes RSP from QT is the measurement strategy which Alice has to employ to accomplish the protocol. In QT Alice needs to perform a complete Bell-state measurement (BSM) between two particles, one from the entangled pair shared with Bob and one carrying the qubit encoding the state to be transferred. For optical implementations, this is a problem because complete BSM is not possible with linear optics only \cite{Calsamiglia01}. In RSP, however, the measurement strategy is more flexible. In the simplest case it may only require projective measurements, as for instance, when Bob knows in advance that Alice will only prepare states lying in a great circle of the Bloch sphere \cite{Pati00}. When the resources for RSP are the same as for QT (1 ebit, 2 cbits), Alice will have to implement a generalized measurement on her qubit. This kind of measurement, which is treated within the formalism of positive operator-valued measurement (POVM), can be implemented through a unitary interaction between the Alice's system and an ancillary one, followed by measurements of both systems \cite{BarnettBook}. The ancillary system may be just a second degree of freedom of Alice's photon and thereby there is no constraint for implementing the POVM with linear optics only \cite{Ahnert05,Croke06}.

So far, several optical realizations of RSP of single qubits have been reported, including pure state of qubits in a superposition of single-photon and vacuum states \cite{Babichev04}, pure and mixed states of photon polarization qubits \cite{Xiang05,Peters05,Liu07,Wu10,Killoran10}, and pure states of spatially encoded photonic qubits \cite{Taguchi08,Kang10}. These realizations were either probabilistic \cite{Babichev04,Xiang05,Peters05,Taguchi08,Kang10} or deterministic \cite{Liu07,Wu10,Killoran10}. In the first case, which some authors refer to RSP just as conditional preparation \cite{Kurucz05}, there is a cost of 1 cbit and the unitary correction by Bob is not required. In the deterministic realizations, \cite{Liu07,Wu10} have successfully implemented the POVM but not the required unitary correction and therefore constituted a proof-of-principle demonstration of the RSP protocol. In Ref.~\cite{Killoran10}, however, the authors have successfully implemented both the POVM and Bob's unitary correction, making the full demonstration of RSP with a cost ranging from 0 cbits for preparation of maximally mixed state to 2 cbits for arbitrary pure states.    

In this article we study the remote preparation of pure qubits states encoded in the discretized one-dimensional transverse momentum of single photons. These so-called spatial qubits are obtained when a paraxial and monochromatic single-photon field is made to pass through a double slit aperture \cite{Neves04}. The advantage offered by this kind of encoding is that one can create high-dimensional qudits simply by increasing the number of slits in the aperture \cite{Neves04,Neves05}. Recent experimental works, employing photon pairs from spontaneous parametric down-conversion, have demonstrated the entanglement between two spatial qubits \cite{Neves07,Taguchi08,Peeters09,Assis11} as well as for two spatial qudits \cite{Neves05,Taguchi09}. Among other applications, spatial qubits and qudits have been used to demonstrate hybrid photonic entanglement \cite{Neves09} and a variety of quantum tomographic techniques as single scan tomography \cite{Taguchi08,Taguchi09} and tomography based on measurements either onto mutually unbiased bases \cite{Lima10,Lima11} or symmetric POVMs \cite{Pimenta10}. Those applications have benefited of the considerable effort for finding methods to measure spatial qubits and qudits \cite{Taguchi08,Neves05,Neves07,Lima08,Taguchi09,Prosser10,Lima11}. 

Here, we assume that Alice and Bob share 1 ebit and we discuss two measurement strategies that she could adopt to remotely prepare his spatial qubit. The first strategy is the well-known spatial postselection, where a single-pixel detector measures the transverse position of the photon between the focal and the image plane of a lens \cite{Neves05,Neves07,Taguchi08,Taguchi09}. The second strategy, somewhat similar to the one proposed by ourselves in \cite{Prosser10}, is a generalized measurement divided into two steps: (i) the implementation of a two-outcome POVM where the photon polarization is used as the ancillary system; (ii) the spatial postselection at the focal plane of the lens by a two-pixel detector in each output of the POVM. Both strategies share the common feature that the ultimate physical property to be measured is the photon transverse position. In this case a realistic description demands us to take into account the finite spatial resolution of the detection system, where the best it can be done is to locate the photon within a narrow interval rather than a point. For both, spatial postselection and generalized measurement, we analyze, \emph{numerically}, the effects of a realistic detector over three figures of merit of the RSP protocol, namely, the probability of preparation, the fidelity and the purity of the remotely prepared states \cite{CommentDet}. It is shown that our strategy improves these figures compared with spatial postselection, at the expense of increasing the classical communication cost as well as the required experimental resources. In addition, we briefly discuss the RSP of spatial qudits and show that by using the generalized measurement strategy we proposed  in \cite{Prosser10} one can prepare arbitrary pure states, unlike spatial postselection alone.

Before proceeding, it is worth to draw a comparison between RSP of spatial and polarization qubits. In the latter, a probabilistic RSP of arbitrary pure states is realized with $50\%$ efficiency through projective measurements \cite{Peters05}. On the other hand, a deterministic implementation employs POVMs and it is $100\%$ efficient (in principle) \cite{Liu07,Wu10,Killoran10}. While it is obvious that the probability of preparation improves, the fidelity and purity of the remotely prepared polarization qubits are limited, in both cases,  only by the precision of the measurement instruments (polarizers, wave plates, etc.) so that they could be one in principle. For spatial qubits, the strategy of spatial postselection would be the analog of projective measurement in polarization. However, here the efficiency is smaller than $50\%$ since the photon propagates in a continuous space confined within the diffraction envelope, and the detector measures only a small fraction of this space. Similarly, our proposed generalized measurement strategy would be the analog of a deterministic implementation for polarization qubits, but, again, due to the diffraction we are not able to get $100\%$ efficiency. Therefore, unlike polarization, for spatial qubits the relation between the figures of merit is not trivially established when one compare the two strategies, since these figures are affected by the diffraction and/or the finite spatial resolution of the detector. Those are the aspects we will approach here.

This article is organized as follows: in Sec.~\ref{sec:RSP} we outline the general framework for RSP of spatial qubits. In Sec.~\ref{sec:spa_post} we discuss the spatial postselection strategy starting with the idealized description of point detectors (Sec.~\ref{sec:spa_post_point}) and then moving on to the realistic description of detectors with finite spatial resolution (Sec.~\ref{sec:det_real}). In the latter case we derive expressions for the three figures of merit to be analyzed and compare our theory with the experimental results from \cite{Taguchi08}. A numerical analysis of the figures of merit is presented in Sec.~\ref{subsec:an1}. Section~\ref{sec:POVM} describes the generalized measurement strategy proposed in this work, first presenting the POVM (Sec.~\ref{subsec:POVM}) and then the detection system. Here, we also begin by considering the idealized point detectors (Sec.~\ref{subsec:2pixpoint}) and next, describe the realistic case of detectors with finite resolution and their effects on RSP (Sec.~\ref{subsec:2pixreal}). We evaluate the figures of merit in this latter case and present our numerical results (Sec.~\ref{subsec:num_an}). In Sec.~\ref{sec:Comparison} we compare the two strategies regarding the figures of merit as well as the resources required to implement the protocol. Section~\ref{sec:RSP_qudits} describes the RSP of spatial qudits either through spatial postselection or the generalized measurement proposed in \cite{Prosser10}, showing the advantages of the latter over the former. Finally, in Sec.~\ref{sec:Conclusion} we conclude the article and briefly discuss a possible extension of our results for the remote preparation of the angular spectrum of a single-photon field as an alternative route for quantum teleportation.

\section{RSP of spatial qubits} \label{sec:RSP}

\begin{figure}
\centerline{\includegraphics[width=0.48\textwidth]{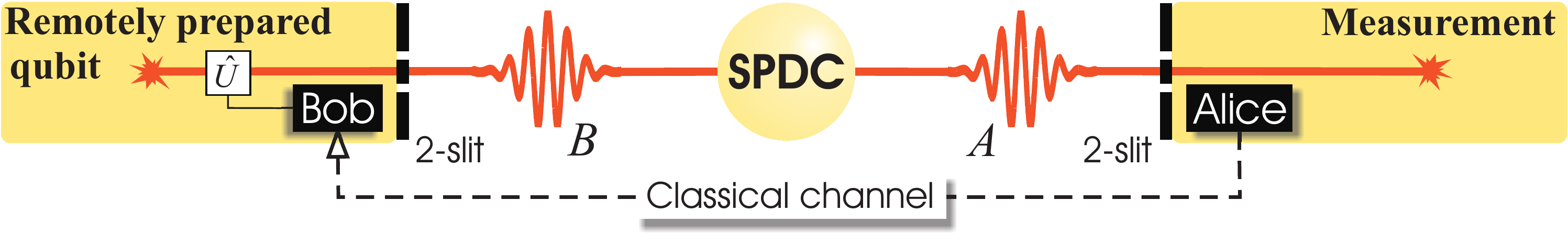}}
\caption{\label{fig:RSP}  (Color online). Schematic representation of RSP of spatial qubits. A SPDC source generates spatially correlated photon pairs whose transverse momentum is discretized by double slits. The two-qubit maximally entangled state (\ref{eq:mes}) is then distributed between Alice and Bob. To remotely prepare the target state (\ref{eq:psi_target}), she measures her photon employing one of the two strategies to be described here and communicates, through a classical channel, her outcome to Bob. Having this information, he may or may not perform a unitary correction, $\hat{U}$, and achieve the desired target state (\ref{eq:psi_target}).   }
\end{figure}

Entangled states of spatial qubits are created by letting each member of the photon pair generated by spontaneous parametric down-conversion (SPDC) cross a double slit aperture as sketched in Fig.~\ref{fig:RSP}. The degree of entanglement between them can be controlled by manipulating the pump beam transverse profile \cite{Neves04,Neves07,Peeters09}. Let us assume that this source creates the maximally entangled state  
\begin{equation}     \label{eq:mes}
|\Psi\rangle_{AB} = \frac{1}{\sqrt{2}}(|l\rangle_{A}|r\rangle_{B}+|r\rangle_{A}|l\rangle_{B}),
\end{equation}
which is shared between Alice ($A$) and Bob ($B$). Here, $|r\rangle$  ($|l\rangle$) denotes the state of the photon transmitted through the right (left) slit. Experimentally, this can be achieved by focusing the pump profile at the double slit plane \cite{Neves04,Neves07}. Now, Alice's goal is to remotely prepare Bob's qubit in the pure state
\begin{equation}     \label{eq:psi_target}
|\psi\rangle_{B} = \alpha|l\rangle + \beta|r\rangle ,
\end{equation}
where $|\alpha|^2+|\beta|^2=1$. For doing so, she measures her spatial qubit through one of the two methods to be discussed here and communicates to Bob, through a classical channel, the outcome of her measurement (see Fig.~\ref{fig:RSP}). With this information, Bob may or may not have to perform a unitary transformation ($\hat{U}$) to recover the target state given by Eq.~(\ref{eq:psi_target}). This, as we will see, depends on the measurement strategy adopted by Alice and the outcome of this measurement.

\section{RSP of spatial qubits through spatial postselection} 
\label{sec:spa_post}

Firstly, let us consider the RSP protocol when Alice remotely prepares the spatial qubit in Bob's laboratory through spatial postselection. This method relies on the detection of the photon transverse position after it has propagated through a given optical system. Typically, this optical system is comprised of a thin convergent lens placed at some distance from the double slit, followed by a detector which can move in the transverse ($x$) and longitudinal ($z$) directions between the focal plane of this lens and the image plane of the double slit \cite{Neves07,Taguchi08,Lima08,Peeters09}. Figure~\ref{fig:RSP_japs}(a) outlines Alice's apparatus to implement spatial postselection. In the following, we discuss two cases regarding Alice's detector system: first, the idealized case of a point detector and second, the realistic case of a single pixel detector with finite resolution.

\subsection{Idealized measurement: The point detector}
\label{sec:spa_post_point}
\subsubsection{Measuring a spatial qubit with a point detector}
A point detector is, obviously, an idealized device which delivers infinite spatial resolution in the measurement of the photon transverse position. Nevertheless, it is important to understand how the RSP protocol would work if Alice had such a device at her disposal, before we move on to the realistic case of detectors with finite resolution in the next subsection. Here, we describe the effect of a measurement by a point detector over an arbitrary state, $\hat{\rho}$, of a spatial qubit. 

\begin{figure}
\centerline{\includegraphics[width=0.48\textwidth]{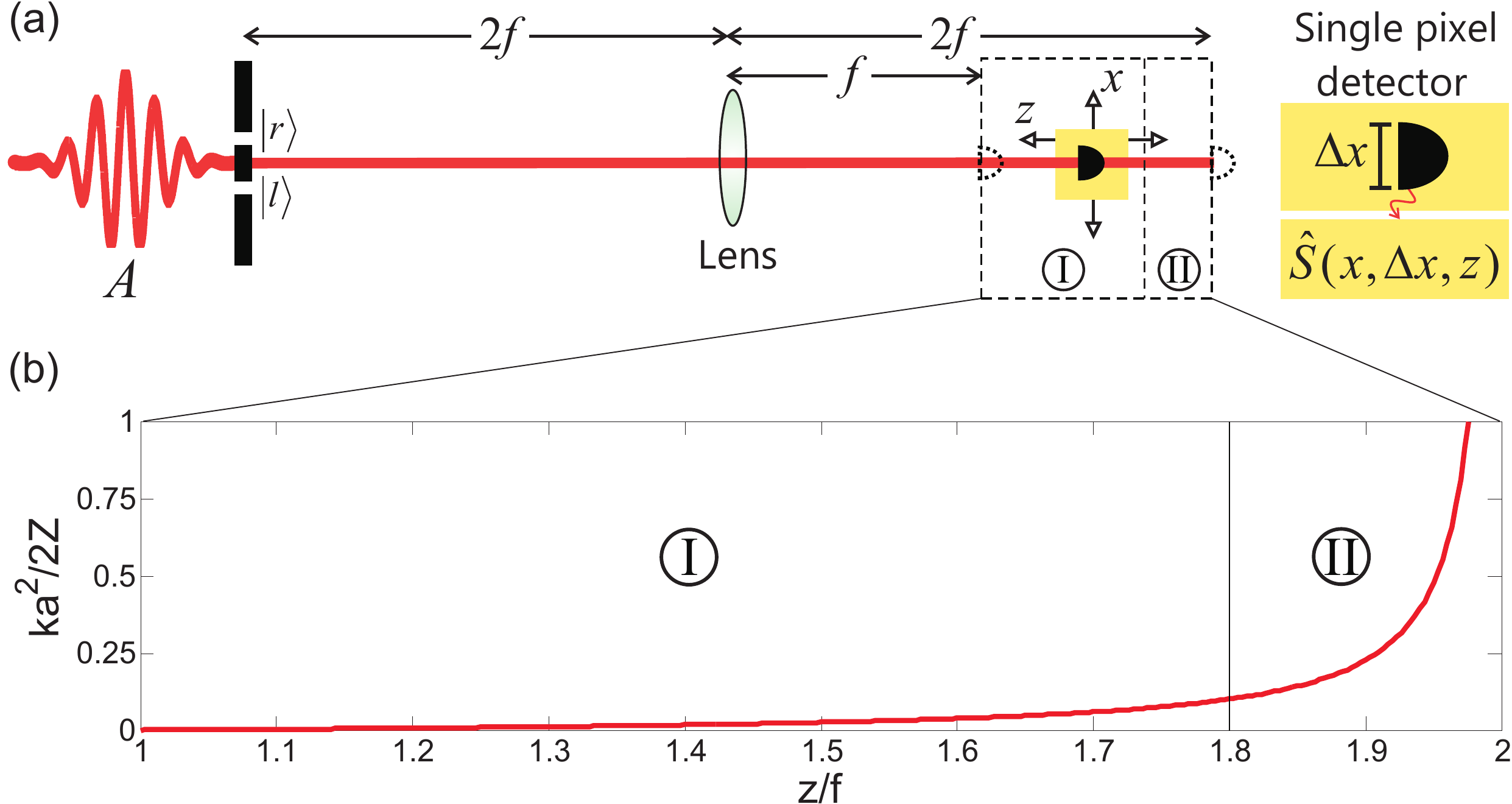}}
\caption{\label{fig:RSP_japs}  (Color online). (a) Schematic representation of the measurement strategy called spatial postselection. A lens of focal length $f$ is placed at a distance $2f$ from the double slit. A single pixel detector is allowed to move between the focal ($z=f$) and the image ($z=2f$) planes in the $x$ and $z$ directions. For a point detector, the corresponding measurement operator is given by Eq.~(\ref{eq:S_projector}). For a detector with finite resolution (as shown in the inset) the measurement operator is given by (\ref{eq:real_op}). The regions I and II set the validity of the Fraunhofer and Fresnel approximations, respectively. The former is valid if the parameter $ka^2/2Z$ is close to zero as shown by the graphic in (b). In the region I the measurement operators are calculated with the amplitudes given by (\ref{eq:Fraunhofer}) while in region II they are calculated from (\ref{eq:Fresnel}).}
\end{figure}

Figure~\ref{fig:RSP_japs}(a) shows that after transmission through the double slit, which discretizes its state, the photon propagates in the continuous space through the optical system. A lens of focal length $f$ is placed at a distance $2f$ from the double slit. Therefore, from our previous discussion, the point detector will be constrained to move, in the transverse direction $x$, between $z=f$ (the focal plane) and $z=2f$ (the plane where the double slit image is formed with a magnification of one). A measurement in a given position $(x,z)$ \emph{postselects} the photon in the (non-normalized) state \cite{Taguchi08}
\begin{equation}     \label{eq:spa_proj}
|\varphi(x,z)\rangle = \varphi_l(x,z)|l\rangle + \varphi_r(x,z)|r\rangle,
\end{equation}
where, up to irrelevant phase factors, the probability amplitudes $\varphi_j(x,z)$ for $j=l,r$ are given by
\begin{eqnarray}
\varphi_j(x,z) & = & \sqrt{\frac{k}{4\pi a\eta Z}}e^{ik\delta_j
x/\eta Z} \nonumber\\
&& \text{}\times\int_{-a}^{a}dx'e^{ikx'(x+\eta\delta_j)/\eta Z} e^{ikx'^2/2Z}.
\label{eq:Fresnel}
\end{eqnarray}
In this expression $a$ is the slit half-width, $\delta_r=d/2$ and $\delta_l=-d/2$, where $d$ is the center-to-center separation between the slits and $k$ is the photon wave number. The parameters $Z=(2f^2-fz)/(z-f)$ and $\eta=(z-f)/f$,
represent the effective longitudinal distance that the photon propagates from the double slit to the detection plane and a scale factor in the detector's transverse position, respectively. Equation~(\ref{eq:Fresnel}) is the Fresnel diffraction integral for a single slit \cite{GoodmanBook} and it is valid everywhere within the range $f\leq z\leq 2f$ where the detector moves. However, it will have analytical solution only if the approximation $Z\gg ka^2/2$ is satisfied. In this case the quadratic phase function in the integral satisfies $e^{ikx'^2/2Z}\approx 1$ and Eq.~(\ref{eq:Fresnel}) reduces to
\begin{equation}      \label{eq:Fraunhofer}
\varphi_j(x,z) =  \sqrt{\frac{ka}{\pi\eta Z}}e^{ik\delta_j
x/\eta Z} \mathrm{sinc}\left[\frac{ka}{\eta Z} (x+\delta_j\eta)\right],
\end{equation}
where $\mathrm{sinc}(\xi)=\sin(\xi)/\xi$. This is the well known Fraunhofer approximation and Eq.~(\ref{eq:Fraunhofer}) yields the Fraunhofer diffraction pattern for a single slit \cite{GoodmanBook}. 

Figure~\ref{fig:RSP_japs}(b) shows how the parameter $ka^2/2Z$ changes between $z=f$ and $z=2f$. This parameter is the maximum value of the argument in the quadratic phase function in Eq.~(\ref{eq:Fresnel}) and determines the regions where the Fraunhofer approximation is valid or not. To plot this curve we used the values of the parameters $a$, $\lambda$ and $f$ shown in Table~\ref{tab:Parameters}. When the approximation is valid we are in the region of Fraunhofer diffraction (region ``I'' in Fig.~\ref{fig:RSP_japs}) and the probability amplitudes $\varphi_j(x,z)$ are calculated from Eq.~(\ref{eq:Fraunhofer}). Otherwise, we are in the region of Fresnel diffraction (region ``II'' in Fig.~\ref{fig:RSP_japs}) and the probability amplitudes are calculated from Eq.~(\ref{eq:Fresnel}). It should be stressed that in both expressions, there is a constant term for each plane $z$ which defines the scale of the diffraction pattern. This term, together with the diffraction envelope itself, has important consequences on the efficiency with which Alice can remotely prepare Bob's spatial qubit. This point will become clear later.

\begin{table}
\begin{center}
\caption{Parameters used in the simulations presented in this work: $a$ and $d$ denote the slits half width and the center-to-center separation between them, respectively, $\lambda$ is the photon wavelength, $f$ is the focal length of the lens, and $\Delta x$ is the detector width.}
\begin{tabular}{ccccccccccccccc}
\hline \hline
\multicolumn{3}{c}{{ Double slit}} &&&&   { Photon}  &&&& { Lens}  &&&&  { Detector}      \\  
$a$ &&  $d$ &&&& $\lambda$ &&&& $f$ &&&& $\Delta x$     \\ \hline
 $40$~$\mu$m && $250$~$\mu$m &&&&  $670$~nm &&&& $30$~cm &&&& $20$~$\mu$m   \\
\hline\hline
\end{tabular}
 \label{tab:Parameters}
\end{center}
\end{table}

From the discussion above the measurement operator associated with the detection by a point detector at the position $(x,z)$ is \cite{Comment_s_op}
\begin{equation}     \label{eq:S_projector}
\hat{s}(x,z) \equiv |\varphi(x,z)\rangle\langle\varphi(x,z)|,
\end{equation}
where $|\varphi(x,z)\rangle$ is given by Eq.~(\ref{eq:spa_proj}). The corresponding detection probability \emph{density} for a spatial qubit in a state $\hat{\rho}$ will be given by (under the assumption that the detector has unit quantum efficiency)
\begin{equation}    \label{eq:dens_prob_A}
\varrho(x,z)=\mathrm{Tr}\left[\hat{s}(x,z)\hat{\rho}\right]. 
\end{equation}

\begin{figure}[t]
\centerline{\rotatebox{0}{\includegraphics[width=0.45\textwidth]{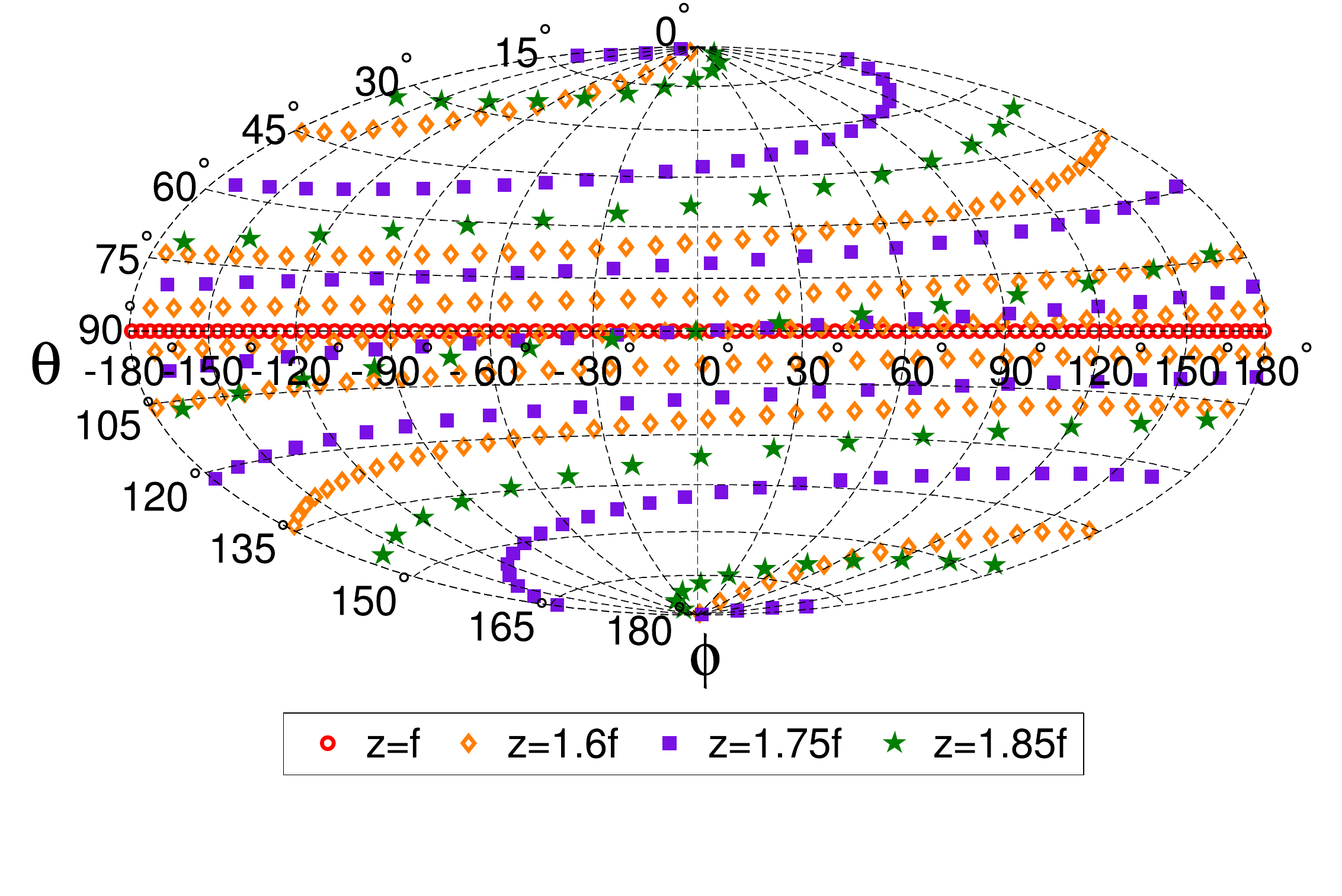}}}
\vspace{-0.8cm}
\caption{\label{fig:Bloch}  (Color online). Bloch sphere (in the Hammer-Aitoff map projection) for the spatial qubits. Each curve corresponds to the trajectory traced out by the state vectors one can measure by displacing the detector in the $x$ direction for the detection planes $z$ shown in the legend. The latitude $\theta\in[0,\pi]$ and the longitude $\phi\in[-\pi,\pi]$ parametrizes an arbitrary pure state $|\psi\rangle=\cos(\theta/2)|l\rangle+e^{i\phi}\sin(\theta/2)|r\rangle$ on the sphere surface.
}
\end{figure}

Identifying each point $(x,z)$ with a normalized vector of the spatial qubit, Taguchi \emph{et al.} \cite{Taguchi08} showed that one can access the whole Bloch sphere of the spatial qubit. Figure~\ref{fig:Bloch} shows some samples of the state vectors one can measure by scanning the point detector in $x$ for distinct longitudinal planes $z$. To calculate the vectors for each point $(x,z)$ we replaced the probability amplitudes given by Eq.~(\ref{eq:Fraunhofer}) into Eq.~(\ref{eq:spa_proj}) and normalize them in order to obtain their polar and azimuthal angles ($\theta$ and $\phi$, respectively) of the Bloch representation. In addition, in the focal plane we considered $x\in[-\pi f/kd,\pi f/kd]$ which suffices for a complete loop around the equator, while for the other planes $z$, $x$ took values which ensure that the trajectories traced out by the vectors were going from one to another pole (see Fig.~\ref{fig:Bloch}) \cite{Comment_xz}. The values of the parameters $a$, $d$, $\lambda$ and $f$  used in this calculation are shown in Table~\ref{tab:Parameters}. As shown by the red circles in Fig.~\ref{fig:Bloch}, at the focal plane ($z=f$) one can measure any state lying only in the equator of the Bloch sphere \cite{Neves07,Lima08}. On the other hand, at the image plane ($z=2f$) one measure only the states $\{|l\rangle,|r\rangle\}$ (not shown in Fig.~\ref{fig:Bloch}) which lie at the poles of the sphere, what is quite intuitive. The interesting aspect noted by Taguchi \emph{et al.} \cite{Taguchi08} is that at intermediate planes ($f<z<2f$) one can measure arbitrary superpositions between $|l\rangle$ and $|r\rangle$. Besides, it is clear from Fig.~\ref{fig:Bloch} that there are several vectors which can be measured at more than one plane. This happens when the trajectories traced out by the vectors for distinct detection planes $z$ intersect each other. For instance, with the exception of the focal plane, the logic basis may be measured at any plane $z\in(f,2f]$. In the same way, with the exception of the image plane, the state $|l\rangle+|r\rangle$ can be measured in $x=0$ for any plane $z\in[f,2f)$. Despite this, there are differences concerning the efficiency with which these ``repeated'' vectors can be measured, due to the diffraction envelope and the scale factor appearing in Eqs.~(\ref{eq:Fresnel}) and (\ref{eq:Fraunhofer}). This means that for an arbitrary state $\hat{\rho}$, if it is possible to postselect it in the same state by measuring in different points $(x,z)$ and $(x',z')$, in general we will have for the detection probability density [Eq.~(\ref{eq:dens_prob_A})]
\begin{equation}
\varrho(x,z)\neq\varrho(x',z').
\end{equation}

\subsubsection{Expected results when Alice uses a point detector}
Let us go back to the RSP protocol of spatial qubits when Alice uses a point detector in the setup of Fig.~\ref{fig:RSP_japs}(a). If photon $A$, which is part of the entangled pair in the state of Eq.~(\ref{eq:mes}), is detected at the position $(x,z)$, then, from Eqs.~(\ref{eq:spa_proj}) and (\ref{eq:S_projector}), the state of Bob's spatial qubit will be given by \begin{equation}       \label{eq:psi_point}
|\psi\rangle_B \propto \varphi_r^*(x,z)|l\rangle+\varphi_l^*(x,z)|r\rangle,
\end{equation}
after she has transmitted 1 cbit to him, namely the outcome of her measurement. Accordingly, if Alice wants to prepare Bob's qubit in the pure state of Eq.~(\ref{eq:psi_target}), then she must place her detector in the position $(x,z)$ such that
\begin{equation}     \label{eq:Alice_condition}
\varphi_l(x,z)\propto\beta^* \hspace{5mm}\mathrm{and}\hspace{5mm}  \varphi_r(x,z)\propto\alpha^*.
\end{equation}
As discussed before, with this procedure she can measure arbitrary pure states so that she can remotely prepare arbitrary pure states as well. After receiving  the cbit from Alice, Bob does not need to perform a unitary correction since, with this strategy, the target state has been already prepared as it was intended to be. From Eq.~(\ref{eq:dens_prob_A}), the probability density for this preparation will be
\begin{eqnarray}
\varrho(x,z) & = & \mathrm{Tr}_{AB}\left\{[\hat{s}(x,z)\otimes\hat{1}_B]|\Psi\rangle_{AB}\langle\Psi| \right\}  \nonumber\\[1mm]
& = & \frac{|\varphi_r(x,z)|^2+|\varphi_l(x,z)|^2}{2},
\label{eq:dens_prob}
\end{eqnarray}
where $\hat{1}_B$ is the identity operator for Bob's qubit, and $\varphi_j(x,z)$ (for $j=l,r$) is given either by Eq.~(\ref{eq:Fresnel}) or (\ref{eq:Fraunhofer}) depending whether Alice measures in region II or I of Fig.~\ref{fig:RSP_japs}, respectively. It is worth mentioning that in this idealized situation, the fidelity and purity of the remotely prepared spatial qubits would be both equal to one.

\subsection{Realistic measurement: Detector with finite resolution}
\label{sec:det_real}

In the real world, any detector has a finite spatial resolution, which means that the best it can do is to locate the photon within some narrow interval $\Delta x$ around the position $x$. This single-pixel detector is illustrated in the inset of Fig.~\ref{fig:RSP_japs}(a). For this device, the associated measurement operators can no longer be written as in Eq.~(\ref{eq:S_projector}). A realistic measurement operator can be derived by noting that the probability of detecting the photon within the range ($x-\Delta x/2,x+\Delta x/2$) in a given transverse plane $z$ is 
\begin{equation}     \label{eq:prob_gen}
P(x,\Delta x,z) = \int_{x-\Delta x/2}^{x+\Delta x/2}\varrho(x',z)\,dx'. \end{equation}
Hence, using Eq.~(\ref{eq:dens_prob_A}) is straightforward to show that
\begin{equation}     \label{eq:prob}
P(x,\Delta x,z) = \mathrm{Tr}\left[\hat{S}(x,\Delta x,z)\hat{\rho}\right],
\end{equation}
where
\begin{equation}      \label{eq:real_op}
\hat{S}(x,\Delta x,z) = \int_{x-\Delta x/2}^{x+\Delta x/2}\hat{s}(x',z)\,dx',
\end{equation}
is the measurement operator for a detector with finite resolution \cite{Comment_S_det}. 

Let us see what are the consequences for the RSP protocol of spatial qubits, when Alice uses this realistic detector in her apparatus. In the same way as before, if she wants to prepare Bob's qubit in the target state given by Eq.~(\ref{eq:psi_target}), she must place her detector at the position $(x,z)$ satisfying the condition (\ref{eq:Alice_condition}). The two-photon state in Eq.~(\ref{eq:mes}) after Alice's measurement is transformed as follows:
\begin{equation}
|\Psi '\rangle_{AB} = \frac{\left\{\left[\hat{S}(x,\Delta x,z)\right]^{1/2}\otimes\hat{1}_B\right\}|\Psi\rangle_{AB}}{\sqrt{_{AB}\langle\Psi|[\hat{S}(x,\Delta x,z)\otimes\hat{1}_B]|\Psi\rangle_{AB}}},
\end{equation}
where the existence of $\left[\hat{S}(x,\Delta x,z)\right]^{1/2}$ is guaranteed by the positivity of $\hat{S}(x,\Delta x,z)$ \cite{Comment_S_det}.
Once she communicates to Bob the outcome of her measurement, his spatial qubit is left in the state
\begin{eqnarray}
\hat{\rho}_B & = & \mathrm{Tr}_A(|\Psi '\rangle_{AB}\langle\Psi '|) \nonumber\\[1mm]
 & = & N
\left(\begin{array}{ccc}
\Phi_{rr}(x,\Delta x,z) && \Phi_{rl}(x,\Delta x,z) \\[1.5mm]
\Phi_{lr}(x,\Delta x,z) && \Phi_{ll}(x,\Delta x,z)
\end{array}\right),
\label{eq:rho_remote}
\end{eqnarray}
where $N=[\Phi_{rr}(x,\Delta x,z)+\Phi_{ll}(x,\Delta x,z)]^{-1}$, is a normalization constant and the matrix elements $\Phi_{ij}(x,\Delta x,z)$ are given by [using Eqs.~(\ref{eq:spa_proj}), (\ref{eq:S_projector}), and (\ref{eq:real_op}) ]
\begin{eqnarray}
\Phi_{ij}(x,\Delta x,z) & \equiv & \langle i|\hat{S}(x,\Delta x,z)|j\rangle \nonumber\\[1mm]
& = & \int_{x-\Delta x/2}^{x-\Delta x/2}  \varphi_i(x',z)\varphi_j^{*}(x',z)\,dx',
\label{eq:Phi_ij}
\end{eqnarray}
for $i,j=r,l$. This representation will be useful to avoid, as much as possible, cumbersome expressions.

As intuitively expected, a detector with finite resolution introduces impurity in the remotely prepared spatial qubit because of the imperfect filtering it provides. Mathematically, this can be seen from Eq.~(\ref{eq:rho_remote}) where the functions $\Phi_{ij}(x,\Delta x,z)$ defined in (\ref{eq:Phi_ij}) cannot factorize in a product of functions like $\varphi_r(x,z)$ and $\varphi_l(x,z)$. This will only happen in the limiting case of a point detector \cite{Taguchi09} where we have
\begin{equation}         \label{eq:lim}
\lim_{\Delta x\rightarrow 0}\frac{\Phi_{ij}(x,\Delta x,z)}{\Delta x}=\varphi_i(x,z)\varphi_j^*(x,z).
\end{equation}
Despite this, the above condition can be approximated as long as the pixel detector width be small compared with the scale where the functions like (\ref{eq:Fresnel}) or (\ref{eq:Fraunhofer}) oscillate \cite{Taguchi09}. At the other extreme, there is a detector with infinite width. In this case, thanks to the orthogonality of the probability amplitudes $\varphi_i(x,z)$ \cite{Comment_s_op} in Eq.~(\ref{eq:Fresnel}), one can show that 
\begin{equation}         \label{eq:lim2}
\lim_{\Delta x\rightarrow \infty}\Phi_{ij}(x,\Delta x,z)=\delta_{ij}.
\end{equation}
In this case, Alice would remotely prepare only the maximally mixed state $\hat{\rho}_B=\frac{1}{2}\hat{1}_B$. This is also an expected result, since a completely opened detector, or a ``bucket'' detector, does not register where the photon has arrived and therefore is equivalent to trace over Alice's photon in Eq.~(\ref{eq:mes}) \cite{Neves07}.

\subsubsection{Figures of merit of the protocol}  \label{sec_realdetector}

To evaluate the performance of the RSP of spatial qubits according to Alice's measurement strategy, we have chosen three figures of merit, namely, the probability of preparation, the fidelity and the purity of the remotely prepared states. 

The probability of preparation is obtained by replacing Eq.~(\ref{eq:dens_prob}) into (\ref{eq:prob_gen}). Using the definition of Eq.~(\ref{eq:Phi_ij}) it can be written as
\begin{equation}   \label{eq:prob_spa_post}
P(x,\Delta x,z)=\frac{\Phi_{rr}(x,\Delta x,z)+\Phi_{ll}(x,\Delta x,z)}{2}.
\end{equation}

To calculate the fidelity between the state remotely prepared by a realistic detector [$\hat{\rho}_B$ in Eq.~(\ref{eq:rho_remote})] and the target state remotely prepared by an idealized point detector [$|\psi\rangle_B$ in Eq.~(\ref{eq:psi_point})], we must normalize the latter and then, using the definition $F=\langle\psi|\hat{\rho}|\psi\rangle$, we get
\begin{equation}     \label{eq:Fidel_spa}
F(x,\Delta x,z) = \frac{\sum_{ij=r,l}\varphi_{i\oplus 1}(x,z)\Phi_{ij}(x,\Delta x,z)\varphi_{j\oplus 1}^{*}(x,z)}{\sum_{mn=r,l}|\varphi_{m}(x,z)|^2\Phi_{nn}(x,\Delta x,z)},
\end{equation} 
where $\oplus$ indicates addition modulo two ($l\oplus 1=r$ and $r\oplus 1=l$).

Finally, the purity of the remotely prepared qubits in Eq.~(\ref{eq:rho_remote}), using the definition $\mathcal{P}=\mathrm{Tr}(\hat{\rho}^2)$, is given by
\begin{equation}        \label{eq:Pur_spa}
\mathcal{P}(x,\Delta x,z) = \frac{\sum_{ij=r,l}|\Phi_{ij}(x,\Delta x,z)|^2}{[\Phi_{rr}(x,\Delta x,z)+\Phi_{ll}(x,\Delta x,z)]^2}.
\end{equation}

With the fidelity and purity, one can quantify the effects of the finite resolution of the detection system over the remotely prepared states. It is easy to check from Eqs.~(\ref{eq:Fidel_spa}) and (\ref{eq:Pur_spa}), respectively, that both quantities are smaller than one for a detector of finite width $\Delta x$. The only exception is for the remote preparation of qubits in the logic basis $\{|l\rangle,|r\rangle\}$ when Alice measures at the image plane with a detector whose width, $\Delta x$, is not large enough to ``see'' both slits simultaneously. In this case, the fidelity of the remotely prepared state and its purity will be one. This happens because when one integrates over the detector width there will be only a change in the phase which is global and thereby, irrelevant. 

In the limiting case of a point detector when Eq.~(\ref{eq:lim}) holds, the fidelity in Eq.~(\ref{eq:Fidel_spa}) and the purity in Eq.~(\ref{eq:Pur_spa}) goes to one for every remotely prepared state, whereas in the case of a bucket detector, the condition (\ref{eq:lim2}) is satisfied and these quantities will be both $1/2$.

\subsubsection{Comparing the expected fidelities with the experimental results by Taguchi \emph{et al.}}

We can compare the theoretical values for the fidelity of the remotely prepared spatial qubits with those obtained in the experiment performed by Taguchi \emph{et al.} in Ref.~\cite{Taguchi08}. In that work, the authors have demonstrated the remote preparation of six states of spatial qubits, through spatial postselection as described in this section. The parameters characterizing their target states [$\alpha$ and $\beta$ of Eq.~(\ref{eq:psi_target})] are shown in the first and second columns of Table~\ref{tab:Fid1}. They were obtained from Alice's detector positions $(x,z)$ provided in \cite{Taguchi08} and the parameters $a$, $d$, $\lambda$ and $f$ they used (shown in the caption of Table~\ref{tab:Fid1}). Replacing these values in Eqs.~(\ref{eq:spa_proj}) and (\ref{eq:Fresnel}) and normalizing the states, we can determine $\alpha$ and $\beta$. Similarly, taking into account the detector width in \cite{Taguchi08} ($\Delta x=20$~$\mu$m) and using Eqs.~(\ref{eq:Fresnel}), (\ref{eq:Phi_ij}), and (\ref{eq:Fidel_spa}), we calculated the expected values of the fidelity between these target states and the remotely prepared ones. The obtained values are shown in the third column of Table~\ref{tab:Fid1}, while the fourth column shows the experimental results of \cite{Taguchi08}. One observes clearly that the experimental values are within the bound imposed by the theoretical ones and furthermore, following their tendency: the states with expected higher fidelities showed higher fidelities (first and sixth rows), those with expected intermediate values showed intermediate fidelities (second and fifth rows), and the same for the states with expected lower fidelities (third and fourth rows). The reduction of the experimentally obtained fidelities, in comparison with the theoretically expected, can be attributed to several factors: the non-perfect preparation of the maximally entangled photon pair, the error in the positioning of Alice's detector and the tomographic process of Bob which also includes a detector with finite resolution. However, even these sources of errors were negligible in \cite{Taguchi08}, the bound on the maximum achievable fidelities would be set by the finite resolution of Alice's detector within the optical system.

\begin{table}
\begin{center}
\caption{Comparison between the theoretical fidelities calculated from Eq.~(\ref{eq:Fidel_spa}) and the experimental ones obtained in Ref.~\cite{Taguchi08}. The first and second columns show the parameters which characterizes the target states [see Eq.~(\ref{eq:psi_target})] intended to be prepared in \cite{Taguchi08}.  The experimental parameters of \cite{Taguchi08} are: $a=20$~$\mu$m, $d=150$~$\mu$m, $\lambda=810$~nm, $f=5$~cm, and $\Delta x=20$~$\mu$m.}
\begin{tabular}{ccccccccc}
\hline \hline
\multicolumn{4}{c}{{\bf Target state} $|\psi\rangle_\mathrm{B}$} &   \multicolumn{4}{c}{\bf Fidelity}      \\  
$|\alpha|$  && $\arg(\beta/\alpha)$ && Theory &&&  Exp. \cite{Taguchi08}    \\ \hline
0.979 && -0.113 &&  98.4\% &&& 88.7\%  \\
0.776 && -2.319 &&  89.9\% &&& 86.1\%  \\
0.742 && -1.159 &&  89.6\% &&& 84.1\%  \\
0.670 && 1.159 &&  89.6\%  &&& 84.1\%  \\
0.631 && 2.319 &&  89.9\%  &&& 87.1\%  \\
0.201 && 0.113 &&  98.4\%  &&& 91.2\%  \\
\hline\hline
\end{tabular}
 \label{tab:Fid1}
\end{center}
\end{table}

\subsection{Numerical analysis}       \label{subsec:an1}

We have analyzed, numerically, the figures of merit of the RSP protocol described so far (see Fig.~\ref{fig:RSP_japs}) and here we present and discuss our results. The parameters used in our simulations are shown in Table~\ref{tab:Parameters} and their values are typically employed in experiments. In particular, the value we have chosen for the detector width, together with the other parameters, allows for a very good spatial filtering without compromising too much the detection rate \cite{Lima10,Lima11}. Our simulations were performed in the following steps:
\begin{enumerate}
\item The surface of the Bloch sphere is divided into a grid of $300\times 600$ pixels $(\theta_i\times\phi_j)$, which give us $1.8\times 10^5$ states to be analyzed. 

\item We choose one figure of merit to maximize. Here this will be either the probability [Eq.~(\ref{eq:prob_spa_post})] or the fidelity [Eq.~(\ref{eq:Fidel_spa})].

\item Alice's detector is scanned in the $x$ direction in steps of 20~$\mu$m ($=\Delta x$) within the range $[-\pi f/k a,\pi f/k a]$ corresponding to the first-order diffraction zone at the focal plane of the lens. Thus for each transverse plane we get $250$ transverse positions $x_m$ ($m=1,\ldots, 250$).

\item This scanning is performed along $3\times 10^5$ transverse planes, $z_n$ ($n=1,\ldots,3\times 10^5$), within the range $[f,2f]$, which gives us a resolution of 1~$\mu$m in the longitudinal direction.

\item For a given plane $z_n$, each point $(x_m,z_n)$ defines an (idealized) remotely prepared state given by Eq.~(\ref{eq:psi_point}) and after normalization we determine its corresponding pixel $(\theta_i,\phi_j)$ in the Bloch sphere. The figure of merit of this state is then calculated and the obtained value is inserted into the pixel $(\theta_i,\phi_j)$. 

\item In the next plane $z_{n+1}$ the procedure is repeated. The blank pixels are occupied with the calculated values of the figure of merit. If some pixel was already occupied from a scan on previous plane(s), then the figure of merit with higher value will override the smaller one. Thus, for a given vector which can be measured in different points, only the one with the highest value of the chosen figure of merit will remain.

\end{enumerate}
This procedure is repeated for all planes $z_n$ so that in the end we are left with the highest values for the figure of merit we intended to maximize. The two remaining figures of merit are calculated at the points where the first one has been maximized. It is important to mention that for $f\leq z_n\leq 1.8f$, that is, in the Fraunhofer approximation [see Fig.~\ref{fig:RSP_japs}(b)], both the state (\ref{eq:psi_point}) and the figures of merit are calculated with the help of the probability amplitude given by Eq.~(\ref{eq:Fraunhofer}), otherwise, for $1.8f< z_n\leq 2f$ we used Eq.~(\ref{eq:Fresnel}).

\begin{figure*}
\centerline{\rotatebox{-0}{\includegraphics[width=1\textwidth]{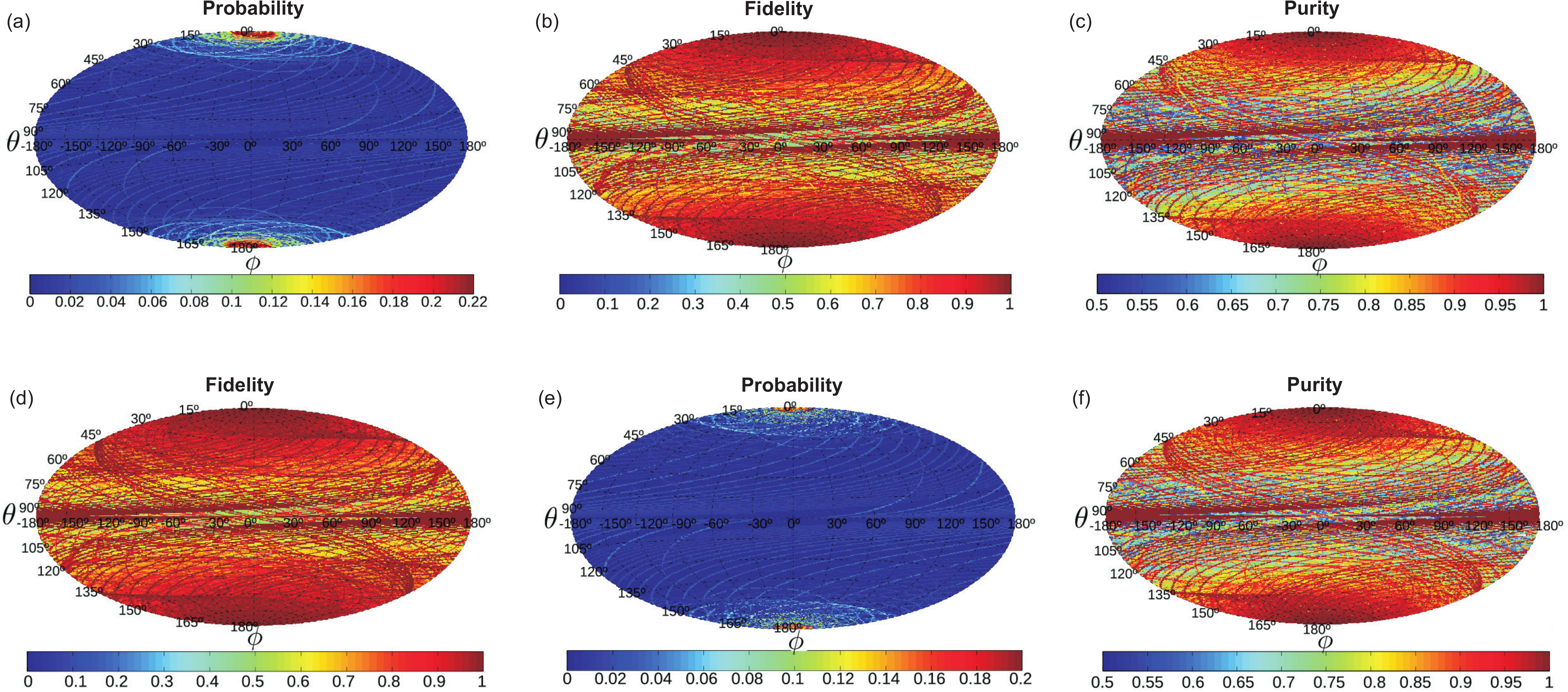}}} \caption{\label{fig:ResultsBloch}  (Color online). Numerical results for the three figures of merit of RSP of spatial qubits through spatial postselection. The first row refers to Alice's choice of maximizing the probability of preparation and the result is shown in the Bloch sphere (a). (b) and (c) show the corresponding fidelity and purity of the remotely prepared states, respectively. In the second row, Alice maximizes the fidelity of the prepared states (d) and their corresponding probability of preparation and purity are shown in (e) and (f), respectively. The procedure for the simulation is described in Sec.~\ref{subsec:an1}.}
\end{figure*}

Firstly, we searched for the higher probabilities of remote preparation and the obtained results are shown in Fig.~\ref{fig:ResultsBloch}(a). The corresponding fidelities and purities of the remotely prepared states are shown in Figs.~\ref{fig:ResultsBloch}(b) and \ref{fig:ResultsBloch}(c), respectively. Secondly, we searched for the higher fidelities of the remotely prepared states and the obtained results are shown in Fig.~\ref{fig:ResultsBloch}(d). Figures~\ref{fig:ResultsBloch}(e) and ~\ref{fig:ResultsBloch}(f) show the corresponding probabilities and purities, respectively. Table~\ref{tab:Comparison} shows the statistics for the three figures of merit taken from the graphics in Figs.~\ref{fig:ResultsBloch}(a)--\ref{fig:ResultsBloch}(c) (second column) and Figs.~\ref{fig:ResultsBloch}(d)--\ref{fig:ResultsBloch}(f) (third column).

\begin{table}
\begin{center}
\caption{Comparison between the average, minimum, maximum values and standard deviation of the probability ($P$), fidelity ($F$) and purity ($\mathcal{P}$).  The second (third) column is associated with RSP via spatial postselection when Alice maximizes the probability (fidelity) and was obtained from Figs.~\ref{fig:ResultsBloch}(a)--\ref{fig:ResultsBloch}(c) [Figs.~\ref{fig:ResultsBloch}(d)--\ref{fig:ResultsBloch}(f)]. The fourth column shows the statistics for RSP via generalized measurement (Sec.~\ref{sec:POVM}) obtained from the results shown in  Figs.~\ref{fig:grafs_POVM}(a)--\ref{fig:grafs_POVM}(c).}
\begin{tabular}{cccccc}
\hline \hline 
 & & \multicolumn{2}{c}{Spatial postselection} & & { POVM}  \\
 & & \emph{Max. probability} & $\;$ \emph{Max. fidelity} & & \\ \hline 	
$\langle P\rangle$ & & $0.02952$ & $0.01871$ & & $0.01547$ \\
$P_{\mathrm{min}}$ & & $0.0001$ & $0.0001$ & & $0.01521$ \\
$P_{\mathrm{max}}$ & & $0.1978$ & $0.1978$ & & $0.01559$ \\
$\sigma_P$ & & $0.04728$ & $0.03310$ & & $0.0001$ \\
 & & & & & \\
$\langle F\rangle$ & & $0.8764$ & $0.9024$ & & $0.9997$ \\
$F_{\mathrm{min}}$ & & $0.1964$ & $0.2046$ & & $0.9995$ \\
$F_{\mathrm{max}}$ & & $1$ & $1$ & & $1$ \\
$\sigma_F$ & & $0.1568$ & $0.1335$ & & $0.0002$ \\
 & & & & & \\
$\langle \mathcal{P}\rangle$ & & $0.8853$ & $0.8987$ & & $0.9995$ \\
$\mathcal{P}_{\mathrm{min}}$ & & $0.5057$ & $0.5058$ & & $0.9989$ \\
$\mathcal{P}_{\mathrm{max}}$ & & $1$ & $1$ & & $1$ \\
$\sigma_\mathcal{P}$ & & $0.1273$ & $0.1183$ & & $0.0004$ \\
\hline\hline
\end{tabular}
 \label{tab:Comparison}
\end{center}
\end{table}

\subsubsection{Discussion}
The effects of diffraction and the scale factor of its associated pattern are clearly observed by the nonuniform probability distribution in either case [Figs.~\ref{fig:ResultsBloch}(a) and \ref{fig:ResultsBloch}(e)]. Also, the effect of a detector with finite spatial resolution can be seen by the non-unity  values for the fidelity [Figs.~\ref{fig:ResultsBloch}(b) and \ref{fig:ResultsBloch}(d)] and purity [Figs.~\ref{fig:ResultsBloch}(c) and \ref{fig:ResultsBloch}(f)] of the remotely prepared states.

When Alice looks for the maximal probabilities, the higher values are found in the poles of the Bloch sphere and at their vicinity. This means that the corresponding states are measured at the image plane and planes close to that, where the light intensity is stronger. For the majority of the superpositions between $|l\rangle$ and $|r\rangle$, however, the higher probabilities are found  far from the image plane so that the achieved values are strongly reduced by diffraction, as can be seen by the blue zones in Fig.~\ref{fig:ResultsBloch}(a). Quantitatively, these results are show in the second column of Table~\ref{tab:Comparison}. 

For several states in the Bloch sphere, the detector position $(x,z)$ where the maximal probabilities were found, are not the same where the fidelity is maximal. This can be seen, by comparing the graphics in the first row with those of the second row of Fig.~\ref{fig:ResultsBloch} or, comparing the second and third column of Table~\ref{tab:Comparison}. In the positions where Alice finds the higher fidelities, the corresponding probabilities of preparation decrease while the purities increase. These results are so because in order to maximize the fidelities, she must place her detector in points where the spatial filtering is stronger, which affects both the probability and purity as described above. The maximum value of one achieved by the fidelity and purity, whatever the figure of merit Alice wants to maximize, is obtained only when she remotely prepares $|l\rangle$ or $|r\rangle$ as we discussed in Sec.~\ref{sec:det_real}.

In our analysis we have considered a detector with fixed width. It would be possible for Alice to use a detector with variable width. This would allow her to improve the  fidelity (or purity) of the remotely prepared states. For instance, let us consider the case where she has maximized the fidelities shown in Fig.~\ref{fig:ResultsBloch}(d). It can be seen there that the lower fidelities are mainly distributed over the regions between the poles and the equator of the Bloch sphere. At the same time, in those regions the probabilities of preparation are, in general, very small. Therefore, reducing the detector width to increase the lower fidelities would, simultaneously, reduce the already low probabilities to prohibitively low levels.

\section{RSP of spatial qubits via generalized measurements}   \label{sec:POVM}

In this section we will present an alternative measurement strategy with which Alice can remotely prepare Bob's spatial qubit. This strategy is based on our recent work \cite{Prosser10} and, as we will see, at the expense of increasing the classical communication cost as well as the experimental resources, it improves the figures of merit of the RSP protocol when compared with spatial postselection alone. The strategy can be divided into two steps: first, the implementation of a two-outcome POVM with the polarization as the ancillary system and second, the spatial postselection at the focal plane of the lens by a two-pixel detector in each output of the POVM. In the following we describe in detail the whole process.

\begin{figure}[t]
\centerline{\includegraphics[width=0.48\textwidth]{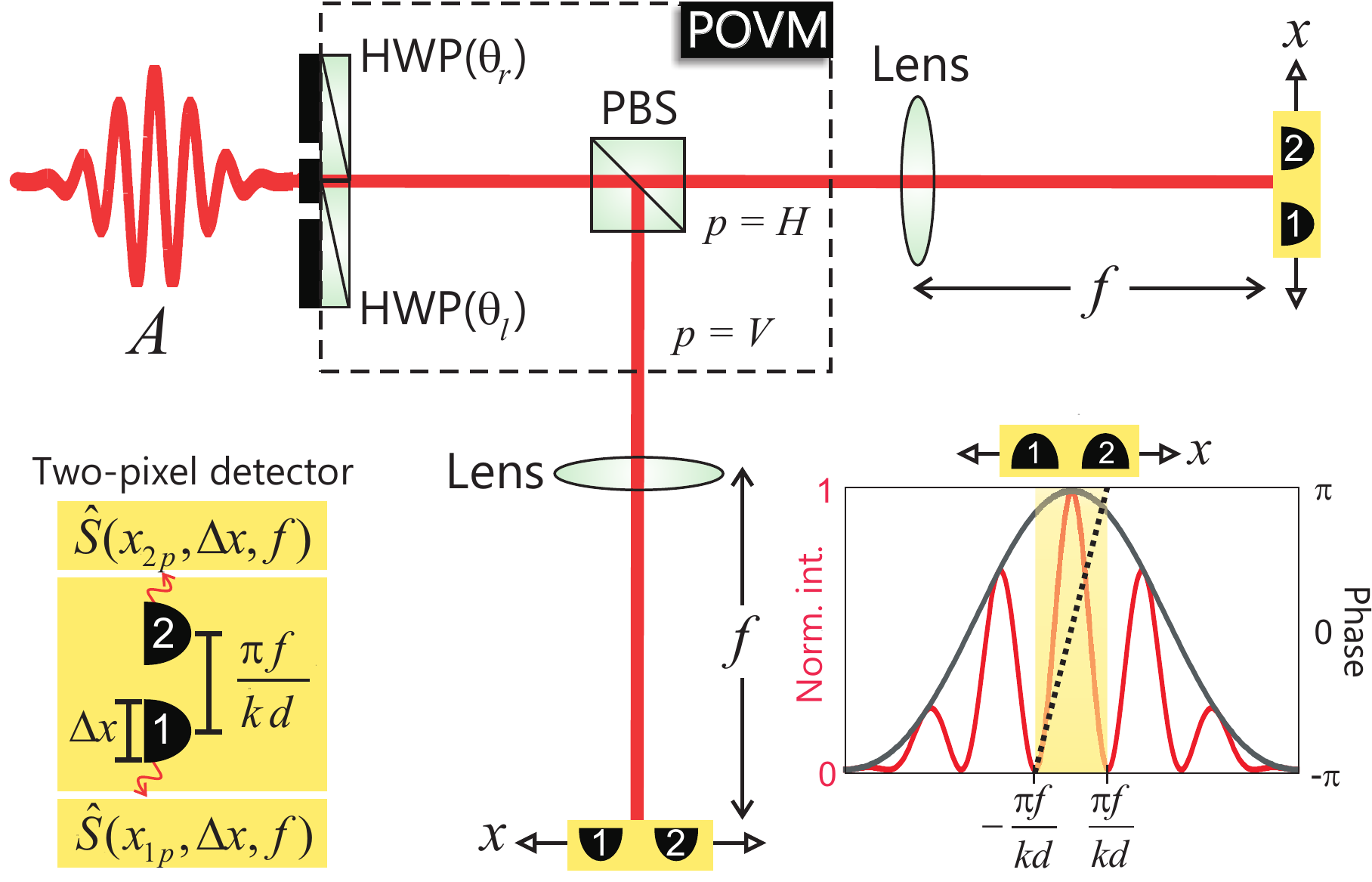}}
\caption{\label{fig:RSP_POVM}  (Color online). Schematic representation of the measurement strategy for RSP of spatial qubits proposed in this work. In the dashed box, a half-wave plate (HWP) behind each slit followed by a polarizing beam-splitter (PBS) implements the two-outcome POVM given by Eq.~(\ref{POVM_el}) \cite{Prosser10}. In each output $p$, a two-pixel detector is placed at the focal plane of a lens and it is allowed to  displace in the $x$ direction. For appropriate choice of the HWP angles [Eq.~(\ref{eq:theta_cond})] and the detector positions [Eqs.~(\ref{eq:det_pos_p}), (\ref{eq:det_pos_2p}) and (\ref{eq:det_chi})], the remotely prepared state is given either by (\ref{eq:1vn})--(\ref{eq:2hn}) for point detectors or by (\ref{eq:rho_jp_B}) for detectors with finite resolution. The left side inset shows a sketch of the two-pixel detector and the fixed separation between the pixels. The right side inset shows the range where the two-pixel detector moves (the shaded region). The dashed line shows how the phase changes according with the detector position.}
\end{figure}

\subsection{The POVM and its physical implementation}    \label{subsec:POVM}

In the first step of the new strategy, Alice will implement a two-outcome POVM on her photon. The physical implementation of a POVM requires the extension of the Hilbert space of the system to be measured, which can be provided by an ancillary quantum system, or \emph{ancilla} \cite{BarnettBook}. Following our proposal \cite{Prosser10}, the ancilla for Alice's spatial qubit will be the photon polarization. For simplicity, let us assume that the overall two-photon state is $|\Psi\rangle_{AB}\otimes|H\rangle_A|H\rangle_B$, where $|\Psi\rangle_{AB}$ is given by Eq.~(\ref{eq:mes}) and $|H\rangle_i$ ($i=A,B$) denotes horizontal polarization. The dashed box in Fig.~\ref{fig:RSP_POVM} sketches the setup for implementing the POVM. Conditional to the passage of photon $A$ through the slit $j$, its polarization is rotated by $2\theta_j$ with a half-wave plate (HWP). This unitary operation entangles the polarization and spatial degrees of freedom. After that, a polarizing beam-splitter (PBS) measures the polarization in the basis $\{|H\rangle,|V\rangle\}$ (where $V$ represents vertical polarization), accomplishing the POVM on the spatial qubit.

Let $\hat{\Pi}_p$ be the elements of the POVM described above, where the subscript $p=H,V$ labels the two possible outcomes. Mathematically, they can be written as
\begin{eqnarray}   \label{POVM_el}
\hat{\Pi}_p & = & \hat{A}_p^\dagger\hat{A}_p \nonumber\\
&=& \langle H|\hat{U}^\dagger|p\rangle\langle p|\hat{U}|H\rangle .
\end{eqnarray}
One can easily check that all properties of a POVM (hermiticity, positivity and completeness) are satisfied. The unitary operator is given by \cite{Comment1}
\begin{equation}  \label{unitary}
\hat{U}=\sum_{j=r,l}|j\rangle\langle j|\otimes\hat{R}(\theta_j),
\end{equation}
where
\begin{equation}  \label{rotation}
\hat{R}(\theta_j)= 
\left(\begin{array}{cc}
\cos(2\theta_j) & -\sin(2\theta_j) \\[2mm]
\sin(2\theta_j) & \cos(2\theta_j)
\end{array}\right)
\end{equation}
describes the conditional polarization rotation. Therefore, when Alice and Bob share the two-spatial qubit state in Eq.~(\ref{eq:mes}) and she implements the POVM (\ref{POVM_el}) on her photon, the two-photon state if the outcome is $p$ will be transformed as follows: 
\begin{eqnarray}
|\Psi_p\rangle_{AB} & = & \frac{1}{\sqrt{P_p}}\left(\hat{A}_p\otimes\hat{1}_B\right)|\Psi\rangle_{AB} \nonumber \\
&=& \frac{1}{2\sqrt{P_p}}\left(\langle p|\theta_l\rangle|l\rangle_A|r\rangle_B+\langle p|\theta_r\rangle|r\rangle_A|l\rangle_B\right), \nonumber \\
\label{eq:Psi_p}
\end{eqnarray}
where 
\begin{equation}      \label{eq:theta}
|\theta_n\rangle=\cos(2\theta_n)|H\rangle+\sin(2\theta_n)|V\rangle,
\end{equation}
for $n=l,r$ and 
\begin{equation}      \label{eq:POVM_prob}
P_p=\frac{|\langle p|\theta_l\rangle|^2+|\langle p|\theta_r\rangle|^2}{2},
\end{equation}
is the probability that photon $A$ exits through the
output $p$, with $P_H+P_V=1$.

The unitary operation of Eq.~(\ref{unitary}) may be experimentally implemented with the help of programmable liquid-crystal displays (LCDs) instead of HWPs behind the slits (Fig.~\ref{fig:RSP_POVM}). LCDs are multipixel optical devices which can, in a controlled way, rotate the light polarization according to the pixel's variable phase retardance. Recently, they have been used to manipulate the amplitude \cite{Pimenta10,Lima09} and/or phase \cite{Lima10,Lima11} of spatial qudits with high accuracy.

\subsection{Measurement at the focal plane by two-pixel point detectors}   \label{subsec:2pixpoint}

\subsubsection{The two-pixel detector}
After being transmitted or reflected at the PBS, photon $A$ propagates toward the detection system which is placed at the focal plane of a lens, as shown in Fig.~\ref{fig:RSP_POVM}. Likewise Sec.~\ref{sec:spa_post}, here we will consider first the idealized situation where Alice has point detectors in her setup, and after that we discuss the realistic case of detectors with finite resolution. Now, rather than a single pixel detector, Alice's detection system (in each output $p=H,V$) will be comprised of two pixels (labeled by 1 and 2) which are mounted over the same platform, so that they can move simultaneously in the transverse direction $x$. In addition, we will impose that the pixels are separated by a fixed transverse distance in order to postselect orthogonal states. Using Eqs.~(\ref{eq:spa_proj}) and (\ref{eq:Fraunhofer}) it is easy to show that this condition is satisfied when
\begin{eqnarray}
\langle\varphi(x_{1p},f)|\varphi(x_{2p},f)\rangle & \propto & \mathrm{sinc}\left(\frac{kax_{1p}}{f}\right)\mathrm{sinc}\left(\frac{kax_{2p}}{f}\right) \nonumber\\
& & \text{}\times\cos\left(\frac{k(x_{2p}-x_{1p})d}{2f}\right) \nonumber \\
& & = 0,
\label{eq:ortog}
\end{eqnarray}
where $x_{jp}$ for $j=1,2$ denotes the transverse position of detector $j$ at the output $p$. On one hand it will be satisfied at the zeros of the sinc functions, which are associated with the zeros of the diffraction envelope, and thus does not concern us. On the other hand, from the cosine function (associated with the interference pattern) we can obtain the fixed distance between the detectors satisfying (\ref{eq:ortog}), which is $x_{2p}-x_{1p}=\pi f/kd$ as shown in the left inset of Fig.~\ref{fig:RSP_POVM}. Due to the periodicity of the cosine function we can further constrain the two-pixel detector to move in a single period within the interval $[-\pi f/kd,\pi f/kd]$ as shown in shaded region of the right inset of Fig.~\ref{fig:RSP_POVM}. As can be seen in this inset, the detection efficiency due to the diffraction envelope is higher in that region. From this discussion, the relation between the detector positions in each arm $p$ can be written as
\begin{equation}      \label{eq:det_pos_p}
x_{2p}=x_{1p}\oplus\frac{\pi f}{kd},
\end{equation}
where $\oplus$ denotes that the addition is performed modulo $2\pi f/kd$, ensuring that both detectors are confined to move in the interval $[-\pi f/kd,\pi f/kd]$, keeping the separation of $\pi f/kd$ as shown in insets of Fig.~\ref{fig:RSP_POVM}.

The measurement operator associated with the point detector $j$ in the focal plane of the lens and at the output $p$ of the PBS (see Fig.~\ref{fig:RSP_POVM}) is given by $\hat{s}(x_{jp},f)$, as defined by Eq.~(\ref{eq:S_projector}). Using Eqs.~(\ref{eq:spa_proj}) and (\ref{eq:Fraunhofer}), the corresponding postselected (non-normalized) state is 
\begin{equation}
|\varphi(x_{jp},f)\rangle=e^{ikdx_{jp}/2f}\varphi_l(x_{jp},f)\left(|l\rangle + e^{ikdx_{jp}/f}|r\rangle\right).
\end{equation} 
Therefore, the two-photon state in Eq.~(\ref{eq:Psi_p}) after Alice's measurement will be transformed as follows:
\begin{eqnarray}
|\Psi_{jp}\rangle_{AB} & \propto & |\varphi(x_{jp},f)\rangle_A|\psi_{jp}\rangle_B,
\end{eqnarray}
where
\begin{equation}  \label{eq:psi_B_jp}
|\psi_{jp}\rangle_B \propto \langle p|\theta_r\rangle|l\rangle+e^{ikdx_{jp}/f}\langle p|\theta_l\rangle|r\rangle,
\end{equation}
is the remotely prepared state of Bob's photon. The vectors $|\theta_n\rangle$ (for $n=r,l$) are given by Eq.~(\ref{eq:theta}). Taking into account the probability for the outcome $p$ [Eq.~(\ref{eq:POVM_prob})] in the POVM, the probability density for this preparation will be
\begin{eqnarray}
\varrho(x_{jp},f) & = & P_p\times\mathrm{Tr}_{AB}\left\{|\Psi_{jp}\rangle_{AB}\langle\Psi_{jp}| \right\}  \nonumber\\[1mm]
& = & \frac{|\langle p|\theta_l\rangle|^2+|\langle p|\theta_r\rangle|^2}{2}|\varphi_l(x_{jp},f)|^2.
\label{eq:dens_prob_p}
\end{eqnarray}

Equation~(\ref{eq:psi_B_jp}) makes clear the roles of the POVM and the detector position for the realization of RSP of spatial qubits. The former will set the real part of the amplitudes $\alpha$ and $\beta$ of the target state (\ref{eq:psi_target}) while the latter, the relative phase $\arg(\beta/\alpha)$.  This will be discussed in more detail next.

\subsubsection{Conditions for remotely preparing the target state}
Having described the POVM and the detection system, we now turn to the conditions that Alice has to accomplish to remotely prepare Bob's photon in the target state given by Eq.~(\ref{eq:psi_target}). Firstly, she must set the wave plates behind the double slit as follows:
\begin{equation}     \label{eq:theta_cond}
\theta_l \equiv \Theta = \frac{1}{2}\cos^{-1}(|\alpha|), \;\;\;\; \theta_r = \frac{\pi}{4}-\Theta.  
\end{equation}
Thus, using Eqs.~(\ref{eq:theta}) and (\ref{eq:psi_B_jp}), the photon $B$ will be prepared in the state 
\begin{eqnarray}
|\psi_{jH}\rangle_B & = & \sin(2\Theta)|l\rangle+e^{ikdx_{jH}/f}\cos(2\Theta)|r\rangle,  \\[2mm]
|\psi_{jV}\rangle_B & = & \cos(2\Theta)|l\rangle+e^{ikdx_{jV}/f}\sin(2\Theta)|r\rangle ,
\end{eqnarray}
given that photon $A$ is measured by the detector $j=1,2$ at the exit port $p=H,V$ of the PBS. In addition, the probability density for the preparation (\ref{eq:dens_prob_p}) becomes
\begin{equation}   \label{eq:p_dens_POVM}
\varrho(x_{jp},f) = \frac{|\varphi_l(x_{jp},f)|^2}{2}.
\end{equation}

The relative phase of the target state, $\arg(\beta/\alpha)$, is controlled by the detector positions. Previously, we saw how the two-pixel detectors move within each arm [see Eq.~(\ref{eq:det_pos_p})]. Now, we impose an extra condition between the positions of these arms ($p=H,V$) given by
\begin{equation}    \label{eq:det_pos_2p}
x_{jH}= - x_{jV},
\end{equation}
such that the whole detection system becomes constrained by Eqs.~(\ref{eq:det_pos_p}) and (\ref{eq:det_pos_2p}). Therefore, the choice of the position of one detector will immediately set the position of the remaining three. Thus, if 
\begin{equation}     \label{eq:det_chi}
x_{1V}=\frac{f}{kd}\left[\arg\left(\frac{\beta}{\alpha}\right)\right],
\end{equation}
we get $x_{1H}=-x_{1V}$ and $x_{2V}=x_{1V}\oplus\pi f/kd=-x_{2H}$.

Accomplishing the conditions (\ref{eq:theta_cond}) and (\ref{eq:det_chi}), Alice will, conditioned to detect her photon with detector $j=1,2$ at output $p=H,V$, remotely prepare Bob's spatial qubit as follows:
\begin{align}
|\psi_{1V}\rangle_B &=\; \alpha|l\rangle + \beta|r\rangle = \;\hat{1}|\psi\rangle_B,\label{eq:1vn}\\[1mm]
|\psi_{2V}\rangle_B &=\; \alpha|l\rangle - \beta|r\rangle = \;\hat\sigma_z|\psi\rangle_B,\\[1mm]
|\psi_{1H}\rangle_B &=\; \beta|l\rangle + \alpha|r\rangle = \;\hat\sigma_x|\psi\rangle_B,\\[1mm]
|\psi_{2H}\rangle_B &=\; \beta|l\rangle - \alpha|r\rangle = \;\hat\sigma_y|\psi\rangle_B,\label{eq:2hn}
\end{align}
where $\hat{\sigma}_n$ (for  $n=x,y,z$) denotes the Pauli operator which Bob must apply on his qubit after receiving the two bits of classical communication from Alice, in order to recover the target state [Eq.~(\ref{eq:psi_target})] she intended to prepare. It is easy to see that with this strategy Alice can prepare arbitrary pure states. First, the POVM which sets the real parts of the amplitudes, allows for values $|\alpha|\in[0,1]$. Second, as shown by the dashed line in the right inset of Fig.~\ref{fig:RSP_POVM}, the range where the detectors move suffices for imparting relative phases from $-\pi$ to $\pi$. As in the case of spatial postselection studied in Sec.~\ref{sec:spa_post_point}, the fidelity and purity of the remotely prepared qubits would be both equal to one, if the detectors were just points.

\subsection{Measurement at the focal plane by two-pixel detectors with finite resolution}     \label{subsec:2pixreal}

Let us now turn to the realistic case where Alice's detection system is composed by detectors with finite resolution. The whole procedure is exactly as we just described and can be summarized as follows:
\begin{enumerate}
\item Alice implements the two-outcome POVM (\ref{POVM_el}) on her half of the maximally entangled photon pair, setting the wave plates satisfying conditions (\ref{eq:theta_cond}).
\item Alice fixes her detection system, in each output $p$, at the focal plane the lens and satisfying conditions (\ref{eq:det_pos_p}), (\ref{eq:det_pos_2p}), and (\ref{eq:det_chi}).
\item Alice communicates 2 bits of classical information to Bob, conditioned to detect her photon with detector $j=1,2$ at output $p=H,V$.
\item Bob performs the appropriate unitary transformation given by (\ref{eq:1vn})--(\ref{eq:2hn})  according with the result communicated by Alice.
\end{enumerate} 

The two-photon state (\ref{eq:mes}) after Alice has implemented the POVM (\ref{POVM_el}) is transformed in (\ref{eq:Psi_p}), whether the outcome was $p$. Thus when she measures photon $A$ with the detector with finite width $\Delta x$, this state is transformed as follows:
\begin{equation}
|\Psi_{jp}'\rangle_{AB} \propto \left\{\left[\hat{S}(x_{jp},\Delta x,f)\right]^{1/2}\otimes\hat{1}_B\right\}|\Psi_p\rangle_{AB},
\end{equation}
where $\hat{S}(x_{jp},\Delta x,f)$ is the measurement operator, defined in Eq.~(\ref{eq:real_op}), for a realistic detector $j$ at the output $p$. Once Alice communicates to Bob the outcome of her measurement, his spatial qubit is left in the state determined by $\hat{\rho}_B^{jp}=\mathrm{Tr}_A\left(|\Psi_{jp}'\rangle_{AB}\langle \Psi_{jp}'|\right)$. Using Eqs.~(\ref{eq:real_op}), (\ref{eq:Phi_ij}), (\ref{eq:theta}) and (\ref{eq:theta_cond}) one can show that
\begin{widetext}
\begin{align}
\hat{\rho}_B^{jp} \propto \begin{cases}
\begin{pmatrix}
\sin^2(2\Theta)\, \Phi_{rr}(x_{jH},\Delta x,f) &  \sin(2\Theta)\cos(2\Theta) \,\Phi_{rl}(x_{jH},\Delta x,f)\\[1mm]
\sin(2\Theta)\cos(2\Theta)\, \Phi_{lr}(x_{jH},\Delta x,f) & \cos^2(2\Theta) \,\Phi_{ll}(x_{jH},\Delta x,f) 
\end{pmatrix}, & \mathrm{if}\;\; p=H,\\
\,\\[1mm]
\begin{pmatrix}
\cos^2(2\Theta)\, \Phi_{rr}(x_{jV},\Delta x,f)&  \sin(2\Theta)\cos(2\Theta)\, \Phi_{rl}(x_{jV},\Delta x,f)\\[1mm]
\sin(2\Theta)\cos(2\Theta)\, \Phi_{lr}(x_{jV},\Delta x,f) & \sin^2(2\Theta)\, \Phi_{ll}(x_{jV},\Delta x,f)
\end{pmatrix}, & \mathrm{if}\;\; p=V.
\end{cases}
\label{eq:rho_jp_B}
\end{align}
\end{widetext}
For point detectors, when the Eq.~(\ref{eq:lim}) holds, $\hat{\rho}_B^{jp}$ reduces to $|\psi_{jp}\rangle_B$ in Eqs.~(\ref{eq:1vn})--(\ref{eq:2hn}). For bucket detectors the condition (\ref{eq:lim2}) holds and Alice would be able to prepare only pure states lying in the poles of the Bloch sphere $\{|l\rangle,|r\rangle\}$, or mixed states along the straight line connecting these poles. This already improves over the spatial postselection in which a bucket detector would prepare just maximally mixed states (see Sec.~\ref{sec:det_real}).

\subsection{Numerical analysis}     \label{subsec:num_an}

Similarly to Sec.~\ref{sec:spa_post}, here we have also analyzed, numerically, the figures of merit of the RSP protocol when it is performed via generalized measurements (see Fig.~\ref{fig:RSP_POVM}). The parameters used in our simulations, shown in Table~\ref{tab:Parameters}, are the same as those used in the previous section.

\begin{figure}[t]
\centerline{\rotatebox{-0}{\includegraphics[width=.52\textwidth]{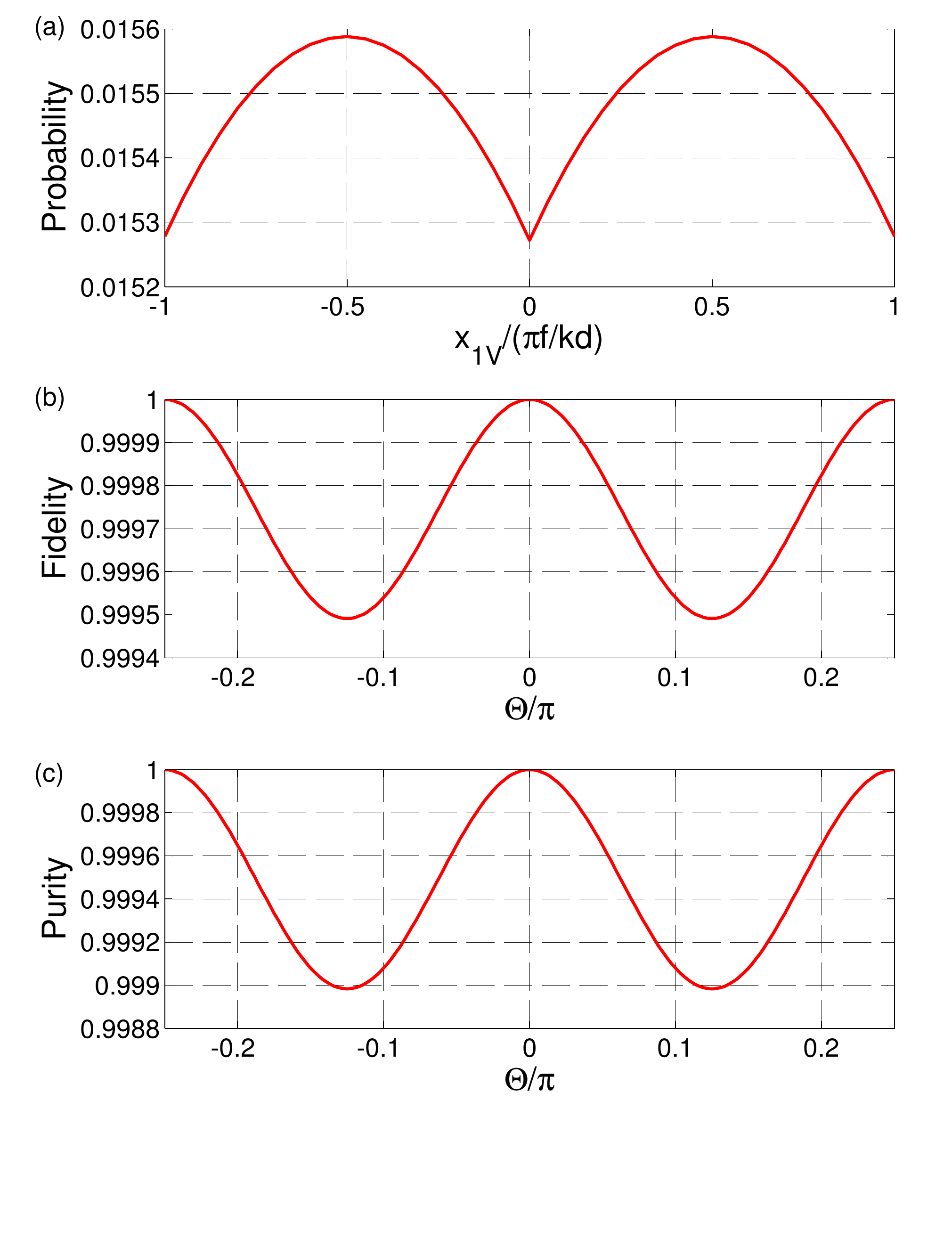}}} 
\vspace{-1.5cm}
\caption{\label{fig:grafs_POVM}  (Color online). Figures of merit for RSP of spatial qubits via generalized measurements (see Fig.~\ref{fig:RSP_POVM}). (a) Total probability of preparation; (b) fidelity and (c) purity of the remotely prepared states. }
\end{figure}

The first figure is the probability of preparation. This is obtained first by integrating the probability density (\ref{eq:p_dens_POVM}) over the detector width $\Delta x$. Using the definition (\ref{eq:Phi_ij}) we get
\begin{equation}
P(x_{jp},\Delta x,f) = \frac{\Phi_{ll}(x_{jp},\Delta x,f)}{2}.
\end{equation}
Since now Alice has four detectors which, upon a click, always prepare Bob's qubit up to a unitary correction, the total probability of preparation will be
\begin{eqnarray}
\nonumber P_{\mathrm{tot}}(x_{1V},\Delta x) & = & \sum_{j=1}^{2}\sum_{p=H,V}P(x_{jp},\Delta x,f)\\
\nonumber & = & \frac{2a}{\lambda f}\int_{x_{1V}-\tfrac{\Delta x}{2}}^{x_{1V}+\tfrac{\Delta x}{2}}dx'\,\Bigg[{\rm sinc}\left(\frac{kax'}{f}\right)\\
& &\text{} + {\rm sinc}\left(\frac{ka}{f}(x'\oplus \pi f/k d)\right)\Bigg].
\end{eqnarray}
Note that the total probability is written as a function of only one detector position, $x_{1V}$, since this sets the position of the others through (\ref{eq:det_pos_p}), (\ref{eq:det_pos_2p}), and (\ref{eq:det_chi}). Figure~\ref{fig:grafs_POVM}(a) shows the plot of this probability which varies around $1\%$ due to the diffraction envelope (see the right inset of Fig.~\ref{fig:RSP_POVM}).

The fidelity of the remotely prepared states must be calculated between the target state prepared when a point detector $j=1,2$ clicks at output $p=H,V$ [$|\psi_{jp}\rangle_{B}$ in Eq.~(\ref{eq:psi_B_jp})] and the state effectively prepared with realistic detectors [$\rho_{B}^{jp}$ in Eq.~(\ref{eq:rho_jp_B})]. Doing so we get
\begin{widetext}
\begin{equation}  \label{eq:fid_povm1}
F(x_{jp},\Delta x,\Theta) = \cos^4(2\Theta) + \sin^4(2\Theta) + \frac{2\cos^2(2\Theta)\sin^2(2\Theta)}{\Phi_{ll}(x_{jp},\Delta x,f)}\displaystyle\int_{x_{j,p}-\Delta x/2}^{x_{j,p}+\Delta x/2}dx'\,|\varphi_l(x',f)|^2\cos\left(\frac{kd(x_{jp}-x')}{f}\right).
\end{equation}
\end{widetext}
The purity of the remotely prepared states $\rho_{B}^{jp}$ in Eq.~(\ref{eq:rho_jp_B}) will be given by
\begin{eqnarray}
\mathcal{P}(x_{jp},\Delta x,\Theta) & = & \cos^4(2\Theta) +\sin^4(2\Theta)\nonumber \\
& & \text{} + \frac{1}{2}\sin^2(4\Theta)\left|\frac{\Phi_{lr}(x_{jp},\Delta x,f)}{\Phi_{ll}(x_{jp},\Delta x,f)}\right|^2. \nonumber\\
&&\label{eq:newPur}
\end{eqnarray}
It turns out that the variation of both quantities, fidelity and purity, within the range $\left[-\frac{\pi f}{kd},\frac{\pi f}{kd}\right]$ where the detectors are allowed to move is negligible. This can be verified by replacing the parameters of Table~\ref{tab:Parameters} into the terms of Eqs.~(\ref{eq:fid_povm1}) and (\ref{eq:newPur}) which depend on $x_{jp}$ and calculating those terms in that range. Therefore the fidelity and purity of the remotely prepared states can be written, respectively, as   
\begin{equation}    \label{eq:fid_approx}
F(\Theta) = 1-0.002[\cos(2\Theta)\sin(2\Theta)]^2,
\end{equation}
\begin{equation}    \label{eq:pur_approx}
\mathcal{P}(\Theta) = 1-0.004[\cos(2\Theta)\sin(2\Theta)]^2,
\end{equation}
and their plots as a function of the HWP angles given by Eq.~(\ref{eq:theta_cond}) are shown in Figs.~\ref{fig:grafs_POVM}(b) and \ref{fig:grafs_POVM}(c). The behavior of both curves can be understood as follows. When $\Theta=\pm\pi/4$ or $\Theta=0$ each output of the POVM completely specify from which slit the photon came. In this case, the states are prepared with unity fidelity and purity no matter the detector width, similarly as the measurement at the image plane in spatial postselection (see Sec.~\ref{sec:spa_post}). On the other hand, for $\Theta=\pm\pi/2$ there is no ``which-slit'' information whatsoever at the outputs of the POVM. Thus, the fidelity and purity depends only on the detector width, reaching their minimum values.  The constant term of Eqs.~(\ref{eq:fid_approx}) and (\ref{eq:pur_approx}) accounts for this dependence. For distinct values of $\Theta$ there is partial which-slit information and interference so that the fidelity and purity took intermediate values.

The statistics for the three figures of merit obtained from the results of Figs.~\ref{fig:grafs_POVM}(a)--(c) is shown in the fourth column of Table~\ref{tab:Comparison}. As it is seen in the graphics of Fig.~\ref{fig:grafs_POVM}, the variation of the three quantities is very small and, accordingly, their standard deviation is negligible. In the next section we discuss these results by comparing them with those obtained from spatial postselection.

\section{Comparison of the methods}     \label{sec:Comparison}

In this section we compare the two measurement strategies for RSP of spatial qubits discussed here, regarding the required resources for each one as well as their figures of merit.

\subsection{Classical communication cost and experimental resources}

Our proposed strategy discussed in Sec.~\ref{sec:POVM} requires 1 ebit and 2 cbits to accomplish the RSP and thereby, increase the classical communication cost of the protocol comparing with spatial postselection (Sec.~\ref{sec:spa_post}) which requires 1 ebit and 1 cbit. Resorting to the comparison between RSP of spatial and polarization qubits established in the Introduction, this is expected since spatial postselection is the analog of a probabilistic implementation in polarization while our strategy is the analog of a deterministic one. 

In terms of experimental resources, our generalized measurement strategy is much more expensive than spatial postselection, in the same way that deterministic RSP of polarization qubits \cite{Killoran10} is more expensive than probabilistic implementations \cite{Peters05}. For spatial postselection, Alice needs a single-pixel detector moving between the focal and image plane, and a logic circuit to send the classical message to Bob. On the other hand, with our proposed strategy Alice needs an apparatus for performing the POVM (which can be done by means of a LCD \cite{Lima11}), four detectors and a more complex logic circuitry to encode the 2 cbits which will trigger Bob's unitary correction [see Eqs.~(\ref{eq:1vn})--(\ref{eq:2hn})]. This last step may be the more challenging in a possible experimental implementation but it is doable in principle. 

\subsection{Figures of merit}

\subsubsection{Probability}
\begin{figure}[t]
\centerline{\rotatebox{-0}{\includegraphics[width=.48\textwidth]{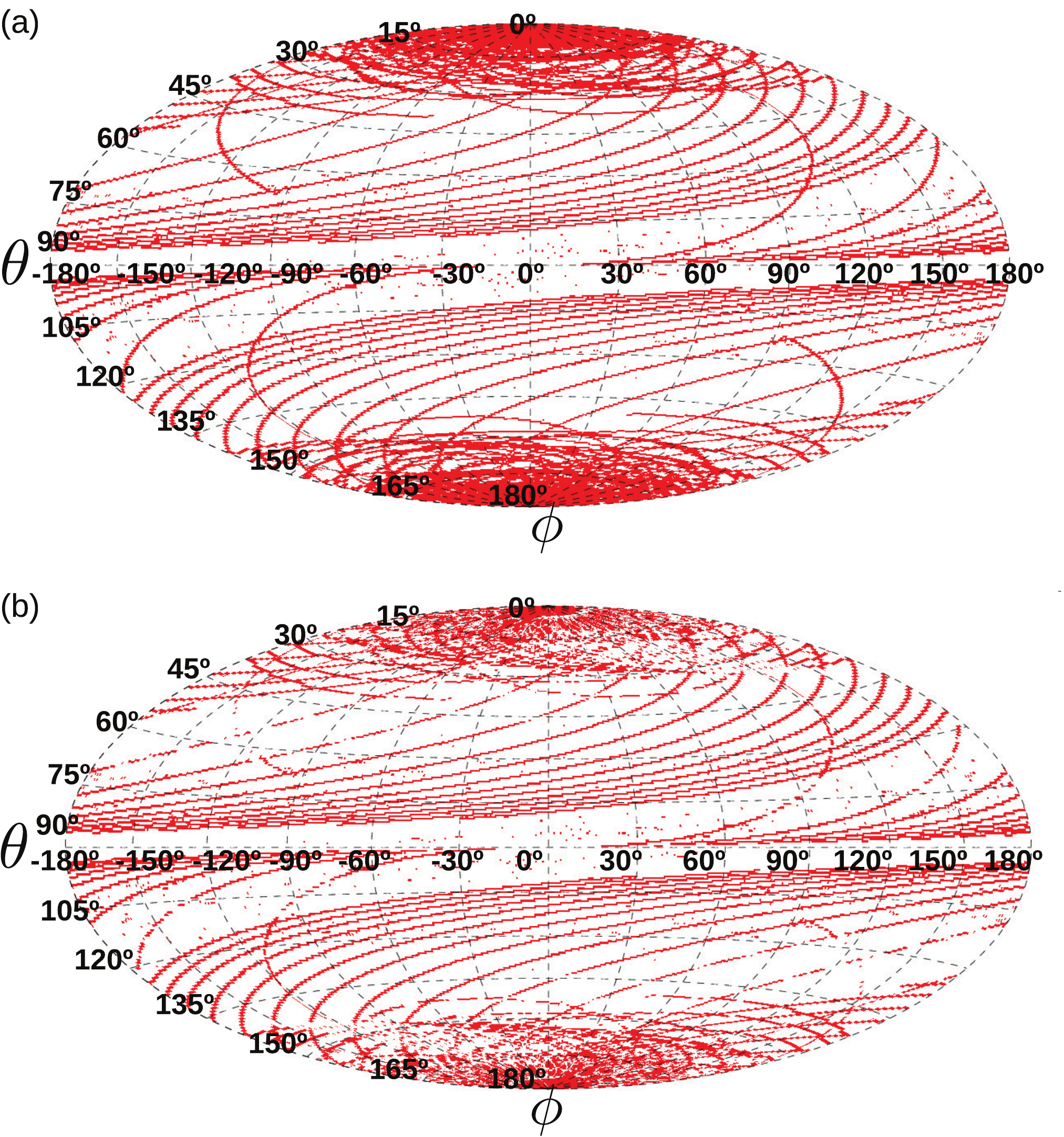}}} \caption{\label{fig:Prob_comparison}  (Color online). Bloch sphere comparing the probabilities for RSP according to Alice's measurement strategy. The red color indicates the states which are prepared with higher probabilities when Alice adopts spatial postselection. In blank, the states prepared with higher probabilities when Alice adopts the generalized measurement. The comparison in (a) refers to Alice's choice of maximizing the probability of preparation [Fig.~\ref{fig:ResultsBloch}(a)] and (b) the fidelity of the remotely prepared states [Fig.~\ref{fig:ResultsBloch}(e)].}
\end{figure}

In Table~\ref{tab:Comparison} it is seen that the average values for the probability of preparation are higher when Alice employs spatial postselection. However, looking for the maximum and minimum values of this probability and the standard deviation of its distribution over the Bloch sphere one can ask how many states are in fact prepared with higher probabilities comparing with our proposed generalized measurement. This comparison is shown in Fig.~\ref{fig:Prob_comparison}, where the red color indicates the states prepared with higher probabilities by spatial postselection, and the blank indicates the states prepared with higher probabilities by generalized measurements. In Fig.~\ref{fig:Prob_comparison}(a)  the comparison refers to Alice's choice of maximizing the probability of preparation via spatial postselection [see Fig.~\ref{fig:ResultsBloch}(a)]. In this case, for $61.4\%$ of the states analyzed, the generalized measurement strategy outperforms spatial postselection. In Fig.~\ref{fig:Prob_comparison}(b)  the comparison refers to Alice's choice of maximizing the fidelity of the remotely prepared states via spatial postselection [the corresponding probability is shown in Fig.~\ref{fig:ResultsBloch}(e)]. In this case, $70.2\%$ of the states analyzed presented higher probabilities when prepared via generalized measurements. Therefore, our measurement strategy provides higher probabilities of remote preparation for the majority of the states when compared with spatial postselection, whatever the figure of merit Alice wants to maximize. 

\subsubsection{Fidelity and purity}

The fidelity and purity of the remotely prepared states are the same for both measurement strategies only when Alice prepares states lying either in the poles or in the equator of the Bloch sphere. Otherwise, the generalized measurement strategy \emph{always} provides higher values for these figures of merit. In fact, in our proposed strategy all the states are prepared with fidelity and purity close to the ideal value of one (achieving one for the preparation of $|l\rangle$ or $|r\rangle$), as can be seen in Figs.~\ref{fig:grafs_POVM}(b) and \ref{fig:grafs_POVM}(c), respectively. On the other hand, for spatial postselection many states are prepared with fidelity and purity much lower than the minimum values achieved with generalized measurement. Quantitatively, this can be seen by comparing the statistics for these quantities shown in Table~\ref{tab:Comparison}.  

As discussed in Sec.~\ref{subsec:an1}, when adopting spatial postselection, Alice could employ a detector with variable width to increase the fidelities (or purities) in the positions which they are low. Doing so, she could, in principle, obtain results as good as those obtained by generalized measurement. However, the required spatial filtering would be so strong that the probability of preparation would become inviable.

\section{RSP of spatial qudits}     \label{sec:RSP_qudits}

 Let us briefly discuss the RSP protocol for spatial qudits. In Ref.~\cite{Taguchi09}, Taguchi \emph{et al.} have experimentally demonstrated the remote preparation of spatial qutrit states using spatial postselection. In this section we discuss the limitations of this method for RSP arising from the increasing of the qudit dimension and, accordingly, the number of free parameters of the state to be prepared. Additionally, we present a generalized measurement \cite{Prosser10} which will overcome these limitations.

To start with, it is well known that a SPDC source can generate maximally entangled states of spatial qudits \cite{Neves04,Neves05}. The setup for this is similar to that shown in Fig.~\ref{fig:RSP} with the double slits replaced by $D$-slits arrays. When the pump beam is focused at the plane of these apertures ($z_\mathrm{ap}$) the two-photon state after transmission through them will be \cite{Neves04,Neves05}
\begin{equation}	   
|\Psi\rangle_{AB}=\frac{1}{\sqrt{D}}\sum_{j=-\ell}^{\ell}e^{i\mu_j}|j\rangle_{A}|-j\rangle_{B},
\label{eq:EntD}
\end{equation}
where $\ell=(D-1)/2$ and $\mu_j=kd^2j^2/2z_\mathrm{ap}$ is a phase factor that depends on fixed experimental parameters and in principle may be compensated by introducing external phase shifts into the slits. Let us assume that this is the case so that in the following we consider $\mu_j=0$ $\forall j$.

Now, Alice's goal is to perform a measurement on her half of the maximally entangled pair, communicate the outcome to Bob, and then remotely prepare his spatial qudit in the pure state 
\begin{equation}   \label{eq:psi_target_qudit}
|\psi\rangle_B = \sum_{j=-\ell}^{\ell}c_j|j\rangle,
\end{equation} 
with $\sum_{j=-\ell}^{\ell}|c_j|^2=1$. Let us again consider the two measurement strategies described in this work, namely, spatial postselection (Sec.~\ref{sec:spa_post}) and generalized measurements (Sec.~\ref{sec:POVM}). Here, however, our discussion will be limited to the idealized case of point detectors.

\subsection{RSP of spatial qudits through spatial postselection}
The setup for spatial postselection of spatial qudits is exactly the same as that sketched in Fig.~\ref{fig:RSP_japs} with the double slit replaced by a $D$-slit array. A point detector constrained to move in the $x$ direction between the focal and image planes of a lens will postselect the state \cite{Taguchi08} 
\begin{equation}     \label{eq:psi_post_qudit}
|\varphi(x,z)\rangle =\sum_{j=-\ell}^{\ell}\varphi_j(x,z)|j\rangle,
\end{equation}
with the probability amplitude $\varphi_j(x,z)$ given either by Eq.~(\ref{eq:Fresnel}) in the Fresnel region or by Eq.~(\ref{eq:Fraunhofer}) in the Fraunhofer region. Thus if Alice wants to prepare Bob's qubit in the target state (\ref{eq:psi_target_qudit}) she must find a position $(x,z)$ to place her detector satisfying
\begin{equation}
\varphi_j(x,z)\propto c^*_{-j}.
\end{equation}

It turns out that while for spatial qubits it is possible to postselect arbitrary pure states through spatial postselection \emph{alone} \cite{Taguchi08}, the same is not true for spatial qudits \cite{ProsserThesis}. This point, which is beyond the scope of the present work, will be addressed in more details elsewhere \cite{Prosser11}. Therefore, when Alice shares the maximally entangled state (\ref{eq:EntD}) with Bob, she cannot prepare arbitrary pure states of his spatial qudit through this strategy.


\subsection{RSP of spatial qudits via generalized measurements}

\begin{figure}[t]
\centerline{\includegraphics[width=0.48\textwidth]{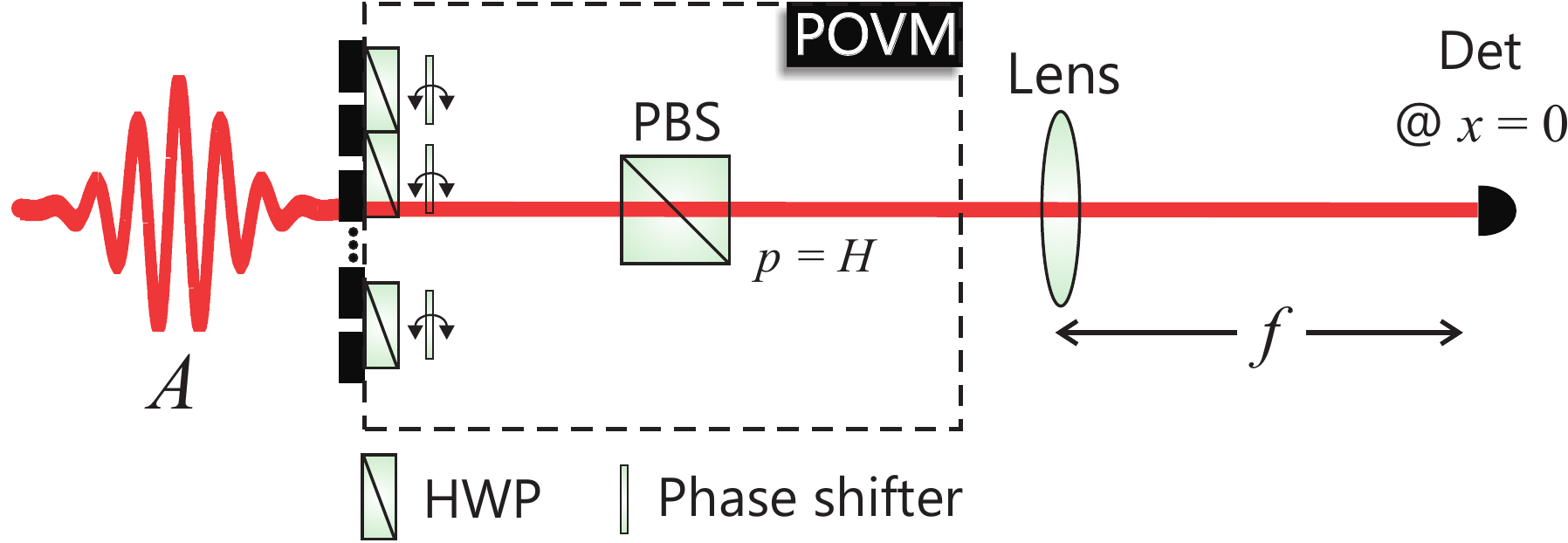}}
\caption{\label{fig:RSP_POVM_qudits}  (Color online). Schematic representation of the measurement strategy for RSP of spatial qudits proposed here.}
\end{figure}
To overcome the difficulties imposed by the spatial postselection for the RSP of spatial qudits, let us assume that Alice  implements the strategy sketched in Fig.~\ref{fig:RSP_POVM_qudits} to measure her half of the maximally entangled pair (\ref{eq:EntD}). Again we assume that the overall two-photon state is, without loss of generality, $|\Psi\rangle_{AB}\otimes|H\rangle_A|H\rangle_B$. Firstly, Alice implements the POVM shown in the dashed box of Fig.~\ref{fig:RSP_POVM_qudits}. Behind each slit there is a HWP followed by a phase shifter. These elements implement a unitary operation which rotates the photon polarization by $2\theta_j$ and adds a phase shift $\phi_j$, conditional to the passage of the photon through the slit $j$ \cite{Prosser10}. Note that this unitary operation differs from  (\ref{unitary}), for spatial qubits, only in the phase shift term. After coupling polarization and spatial degrees of freedom, the former is measured by a PBS. Secondly, Alice proceed with the measurement only if the outcome of the POVM was $p=H$; otherwise she does nothing. In case of success, the photon is detected by a single-pixel point detector in a \emph{fixed} transverse position, $x=0$, at the focal plane of a lens, $z=f$. From Eqs.~(\ref{eq:Fraunhofer}) and (\ref{eq:psi_post_qudit}) one can check that this detection postselects the photon in a state proportional to $\sum_j|j\rangle$. Therefore, after communicating the outcome of her measurement to Bob, his spatial qudit is left in the state 
\begin{equation}
|\psi\rangle_B=\sum_{j=-\ell}^{\ell}e^{i\phi_j}\cos(2\theta_j)|-j\rangle.
\end{equation}
In order to prepare the target state given by Eq.~(\ref{eq:psi_target_qudit}), she must set her HWPs and phase shifters, respectively, as follows:
\begin{equation}
\theta_{j}=\frac{1}{2}\cos^{-1}(|c_{-j}|),\hspace{0.5cm}\phi_{j}=\arg(c_{-j}).
\label{conditionsdimD}
\end{equation}
With the conditions (\ref{conditionsdimD}), the outcome $p=H$ of the POVM implemented on Alice's photon occurs with probability $1/D$ for every setting of the plates. 

This strategy for RSP of spatial qudits differs from that for qubits (see Fig.~\ref{fig:RSP_POVM}) in three aspects: (i) the already mentioned unitary operation; (ii) just one output of the POVM is taken into account since for qudits the second output is useless \cite{Prosser10}; (iii) it employs a single-pixel detector in a fixed transverse position at the focal plane. In this sense, this strategy is comparable to spatial postselection in terms of the classical communication cost. However, RSP of spatial qudits via generalized measurement, as proposed here, offers major advantages from the practical point of view: it does not require the displacement of the detection system and may be experimentally implemented with programmable LCDs \cite{Lima09,Lima10,Lima11,Pimenta10}. This would result in a considerable reduction in experiment time, since instead of positioning the detector for each measurement, we would have only to update the predefined configuration (\ref{conditionsdimD}) in the LCDs and measure the photon at the focal plane. More importantly, our strategy allows for the remote preparation of arbitrary pure states of spatial qudits which is not possible with spatial postselection alone.

\section{Conclusion}       \label{sec:Conclusion}

We have investigated, theoretically, the quantum communication protocol of RSP for pure states of spatial qubits. Two measurement strategies were studied, spatial postselection and a generalized measurement proposed by ourselves. In both cases we have analyzed the diffraction effects as well as the finite spatial resolution of the detection system over three figures of merit of the protocol, namely probability of preparation, fidelity, and purity of the remotely prepared state. It was shown that our strategy improves those three quantities compared with spatial postselection: it provides higher probabilities of preparation for the majority of the states in the Bloch sphere and allows their preparation with fidelities and purities close to the ideal value of one. Additionally, we have presented a scheme for RSP of spatial qudits based on our measurement strategy proposed in \cite{Prosser10}, which allows for the preparation of arbitrary pure states unlike spatial postselection alone \cite{ProsserThesis,Prosser11}. In our schemes the crucial step in the realization of the protocol, is the implementation of the POVM with the polarization as the ancillary system. Experimentally, this could be achieved with programmable LCDs as recently demonstrated \cite{Lima09,Lima11,Pimenta10}.

A possible extension of our work is the investigation of RSP of the angular spectrum of a single-photon field. This continuous degree of freedom allows one to encode a large amount of information. RSP could be a reliable method for transmitting this information. Recently, Walborn \emph{et al.} \cite{Walborn07} proposed a QT scheme to accomplish such task for an unknown angular spectrum. In their scheme, Alice and Bob share a two-photon entangled state produced by SPDC and a third photon, carrying the angular spectrum to be transmitted, is in possession of Alice. Then, she performs a joint measurement on her photons by injecting them into a nonlinear up-conversion crystal cut for second harmonic generation and send the outcome to Bob through a classical channel. Depending on the outcome, Bob performs a unitary correction to recover the initial state. Nevertheless, there are two liming factors for the realization of the protocol. First, the state produced by SPDC is not equivalent to the maximally entangled Einstein-Podolsky-Rosen state \cite{Einstein35} (which is required for faithful teleportation) and thereby, the fidelity of the transmitted state is reduced. Second, the required nonlinear interaction in Alice's measurement strategy is very inefficient ($\sim 10^{-7}$), which compromises the overall efficiency of the scheme. If Alice had knowledge of the angular spectrum to be transmitted, RSP would be the better choice for this task, since the above difficulties imposed by QT would be circumvented. First, because RSP could be faithfully performed with nonmaximally entangled states \cite{Ye04} and second, the measurement strategy employed by Alice would not require nonlinear interactions.

\section*{ACKNOWLEDGMENTS}
This work was supported by Grant Nos. CONICYT PFB08-024, Milenio ICM NC10-030-F, PBCT Red21 and FONDECYT 11085057. M. A. S. P. acknowledges the financial support from CONICYT.


\begin{thebibliography}{99}
\bibitem{Comment_xz}  For positions $x$ beyond those considered, one can still measure other states. However, the efficiency of the measurement would be strongly affected by diffraction effects and therefore we have not considered such cases.
\bibitem{Comment_S_det} The realistic measurement operators $\hat{S}(x,\Delta x,z)$ in (\ref{eq:real_op}) also constitutes POVM elements in the infinitely discrete space which is divided into pixels of width $\Delta x$. Again, the hermiticity and positivity of $\hat{S}(x,\Delta x,z)$ follow directly from its definition. The completeness relation, for an arbitrary plane $z$, can be demonstrated by dividing the transverse plane in segments labeled by $j$, each one with width $\Delta x$ and centered on $x_j$. For each interval there is a corresponding measurement operator  $\hat{S}(x_j,\Delta x,z)$ for which is straightforward to show (using the completeness relation for $\hat{s}(x,z)$ \cite{Comment_s_op}) $\sum_{j=-\infty}^{\infty}\hat{S}(x_j,\Delta x,z)=\hat{1}_2$.
\bibitem{Comment1} In Ref.~\cite{Prosser10} we have considered a unitary operation that not only rotate the photon polarization but also added phase shifts at each slit. For RSP of spatial qubits this procedure for introducing relative phases would not be suitable to accomplish the protocol and we transferred this task for the two-pixel detector positions at the focal plane.
\end{thebibliography}
\end{document}